\documentclass[aps,amssymb,amsfonts,amsmath,a4paper,article]{revtex4}
\usepackage{longtable}

\newif\ifpdf
\ifx\pdfoutput\undefined
  \pdffalse
\else
  \pdfoutput-1
  \pdftrue
\fi
\ifpdf
     \usepackage[pdftex]{graphicx}
     \DeclareGraphicsExtensions{.pdf}
\else
     \usepackage[dvips]{graphicx}
     \DeclareGraphicsExtensions{.eps}
\fi

\makeatletter
\renewcommand{\@seccntformat}[1]{\large \csname the#1\endcsname .
\hspace{0.5em}}
\makeatother
\numberwithin{equation}{section}

\begin{document}

\title{Fluctuation-Response Relation and modeling in systems with fast and slow dynamics}
\author{G. Lacorata$^1$ and A. Vulpiani$^2$}
\address{$^1$Institute of Atmospheric and Climate Sciences, National Research Council, 
Str. Monteroni, I-73100 Lecce, Italy.}
\address{$^2$Department of Physics and INFN, University of Rome "La Sapienza", P.le A. Moro 2, I-00185 Roma, Italy.}

%

\begin{abstract}
We show how a general formulation of the Fluctuation-Response Relation is able to describe in detail the 
connection between response properties to external perturbations and spontaneous fluctuations   
in systems with fast and slow variables. The method is tested by using the 360-variable Lorenz-96 
model, where slow and fast variables are coupled to one another with reciprocal feedback, and a simplified low 
dimensional system. In the Fluctuation-Response context, the influence of the fast dynamics on the slow dynamics 
relies in a non trivial behavior of a suitable quadratic response function. This has important consequences 
for the modeling of the slow dynamics in terms of a Langevin equation: beyond a certain intrinsic time  interval 
even the optimal model can give just statistical prediction. 
\end{abstract}

\maketitle

\section{Introduction}

One important aspect of climate dynamics is 
the study of  the response to perturbations of the external
forcings,  or of the control parameters.
In very general terms, let us consider the symbolic evolution equation:
\begin{equation}
\label{I.1}
{{d{\bf X}} \over {dt}}= {\bf Q}({\bf X})
\end{equation}
where ${\bf X}$ is the state vector for the system, and ${\bf Q}({\bf X})$
represents complicated dynamical processes. 
As far as climate modeling is concerned, one of the most interesting properties 
to study is the so-called
Fluctuation-Response relation (FRR), i.e.
 the possibility, at least in principle, 
to understand the behavior of the system (\ref{I.1}) under 
perturbations (e.g. a volcanic eruption, 
or a change of the $C \, O_2$ concentration)  in terms of
the knowledge obtained from its past  time history
(Leith 1975, 1978, Dymnikov and Gritsoun, 2001).

The average effect on the variable $X_i(t)$ of an infinitesimal
 perturbation $\delta {\bf f}(t)$ in (\ref{I.1}),
i.e. ${\bf Q}({\bf X}) \to {\bf Q}({\bf X}) +\delta{\bf f}(t)$,
 can be written in  terms of the response matrix $R_{ij}(t)$.
 If $\delta {\bf f}(t)=0$ for $t<0$ one has:
\begin{equation}
\label{I.2}
\overline{\delta X_{i}(t) }=\sum_{j} \int_0^t 
 R_{ij}(t-t')  \delta  f_{j}(t') dt' \, 
\end{equation}
where  $R_{ij}(t)$ is the average response 
of the variable $X_i$  at time $t$ with respect to a perturbation of $X_j$
at time $0$.

The basic point  is, of course, how to express $R_{ij}(t)$ in terms
of correlation functions of the unperturbed system. 
The answer to this problem is the issue of
the  Fluctuation-Response theory.
This field has been initially 
developed  in the context of equilibrium statistical 
mechanics of Hamiltonian systems; this 
generated some confusion  and misleading ideas
on its validity. As a matter of fact, it is possible to show that a generalized FRR holds 
under rather general hypotheses (Deker and Haake, 1975, Falcioni et al., 1990): 
the FRR is also valid in non Hamiltonian systems.  
It is interesting to note that, although
stochastic and deterministic systems, from
a conceptual (and technical) point of view, are somehow 
rather different, the same FRR holds in both cases,
see Appendix A. For this reason, in the following, we will not
separate the two cases. In addition, a FRR holds 
also for not infinitesimal perturbation (Boffetta et al., 2003). 
From many aspects, the FRR issues in climate systems are rather similar
to those in fluids dynamics: we have to deal with  non Hamiltonian 
and non linear systems whose invariant measure is non  Gaussian (Kraichnan, 2000). 
On the other hand, it is obviously impossible
to model climate dynamics with equations obtained from first principles, so typically
it is necessary to work with simple raw models or just to deal 
with experimental signals (Ditlevsen, 1999, Marwan et al., 2003).
In addition, in climate problems (and more in general in Geophysics)
the study of infinitesimal perturbation is rather academic,
while a much more interesting question is the relaxation of large
perturbations in the system due to fast changes of the parameters.  
Numerical simulations show that, in systems 
with one single time scale (e.g. low dimensional chaotic model
as the Lorenz one), 
the amplitude of the perturbation is not so important, 
(see Appendix A, and Boffetta et al., 2003). 
On the contrary, in the case of different characteristic times, 
the amplitude of the perturbation  
can play a major role in determining the response, 
because different amplitudes may affect features 
with different time properties (Boffetta et al., 2003). 
Starting from the seminal works of Leith (1975, 1978),
who proposed the use of FRR for the
response of the climatic system to changes in the external
forcing, many authors tried to apply this relation to
different geophysical problems, ranging from simplified models 
(Bell, 1980), to general circulation models (North et al., 1999, 
Cionni et al., 2004) 
and to the covariance of satellite radiance spectra 
(Haskins et al., 1999). 
For recent works on the application of the FRR to the sensitivity
problem and the predictability see Gritsoun and Dymnikov (1999),
Gritsoun (2001), Gritsoun et al. (2002), Dymnikov and  Gritsoun (2005),
Dymnikov (2004), Abramov and Majda (2007), and Gritsoun and Branstator (2007). 
In most works, the FRR has been invoked in its
Gaussian version, see below,  which has been used
as a kind of approximation, often without a precise idea of its limits
of applicability. 
In principle, according to Lorenz (1996), one has to consider two kinds
of sensitivity: to the initial conditions (first kind) and to
the parameters (e.g. external forcing) of the system (second kind). 
On the other hand, if one considers just infinitesimal perturbations,
it is possible to describe the second kind problem in terms of the first
one. Unfortunately, this is not true for non infinitesimal perturbations.

In this paper we study, in the FRR framework, systems with more than one characteristic time. 
In section 2 we recall the theoretical basis of the FRR issue. 
In section 3 we describe the analysis we have performed on two dynamical systems. 
The first one, is a model introduced by Lorenz (1996), which contains some of the 
relevant features of climate systems, i.e. the presence of fast and slow variables (see Fraedrich, 2003, for 
a discussion about short and long-term properties of complex multiscale systems like the atmosphere).  
We consider, at this regard, the problem of the parameterization of the fast variables via a suitable 
renormalization of the parameters appearing in the slow dynamics equations, and the addition of a random forcing. 
The second one is a very simplified system consisting, basically, of a slow variable which 
fluctuates around two states, coupled to fast chaotic variables. 
The specific structure of this system suggests a modeling of 
the slow variable in terms of a stochastic differential equation. 
We will see how, even in absence of a Gaussian statistics, the correlation functions of the slow (fast) 
variables have, at least, a qualitative resemblance with response functions to perturbations on the 
slow (fast) degrees of freedom. In addition, although the average response of a slow variable to perturbations 
of the fast components is zero, the  influence of the fast dynamics on the slow dynamics cannot be neglected. 
This fact is well highlighted by a non trivial behavior of a suitable quadratic response function 
(Hohenberg and Shraiman, 1989). 
In the framework of the complexity in random dynamical systems, one has to deal with a similar behavior: 
the relevant `complexity' of the system is obtained by considering the divergence of nearby 
trajectories evolving with two different noise realizations (Paladin et al., 1995). 
This has important consequences for the modeling of the slow dynamics in terms of a Langevin equation: 
beyond a certain intrinsic time interval (determined by the shape of the quadratic response function) 
even the optimal model can give just statistical predictions (for general discussion about the skills and the limits 
of predictability of climatic models see Cane, 2003).  
The conclusions and the discussion of the results obtained in this work are contained in section 4, while  
the Appendices are devoted to some technical aspects.  

\section{Theoretical background on FRR}
\label{sec:TheoreticalBackgroundOnFRR}

For the sake of completeness we summarize here some basic results regarding the FRR 
(see Appendix A for technical details).  
Let us consider a dynamical system $ {\bf X}(0) \to {\bf X}(t)=U^t {\bf X}(0)$
whose time evolution can even be not
completely deterministic (e.g. stochastic differential equations),  
with states ${\bf X}$ belonging to a $N$-dimensional vector space.  
We assume: a) the existence of an invariant probability 
distribution $\rho({\bf X})$, for which some ``absolute continuity'' 
type conditions are required (see Appendix A);
 b) the mixing\footnote{A dynamical system is mixing if, for $t \to \infty$, 
 $\langle f(U^t{\bf X})g({\bf X}) \rangle \to \langle f({\bf X}) \rangle \langle g({\bf X}) \rangle$,  
 where the average is over the invariant probability distribution and 
 $f$ and $g$ are $L_2$ functions.} 
 character of the system (from which its
ergodicity follows). 

At time $t=0$ we introduce a perturbation $\delta {\bf X}(0)$ on the variable
${\bf X}(0)$. 
For the quantity $\overline{\delta X_i}(t)$, in the case of
an infinitesimal perturbation $\delta {\bf X}(0) = (\delta X_1(0)
\cdots \delta X_N(0))$ one obtains:
\begin{equation}
\label{I.3}
\overline{\delta X_i} \, (t)=
\sum_j R_{ij}(t) \delta X_j(0)
\end{equation}
where  the linear response functions (according to FRR) are 
\begin{equation}
\label{I.4}
R_{ij}(t) = - \Biggl \langle X_i(t) \left.
 \frac{\partial \ln \rho({\bf X})} {\partial X_j} \right|_{t=0}
\Biggr  \rangle \, . 
\end{equation} 
In the following $\langle () \rangle$ indicates the average on the unperturbed system, while 
$\overline{()}$ indicates the mean value of perturbed quantities. 
The operating definition of  $R_{ij}(t)$ in numerical simulations 
is the following. We perturbe the variable $X_j$ at time $t=t_0$ with a small
perturbation of amplitude $\delta X_j(0)$ and then evaluate the 
separation component $\delta X_i(t|t_0)$ between the two trajectories 
${\bf X}(t)$ and ${\bf X}'(t)$ which are integrated up to a
 prescribed time $t_1=t_0+\Delta t$.
 At time $t=t_1$, the variable $X_j$ of the
reference trajectory is again perturbed with the same $\delta X_j(0)$,
and a new sample $\delta {\bf X}(t|t_1)$ is computed and so forth.  
The procedure is repeated $M \gg 1$ times and the mean response is then 
given by: 
$$
R_{ij}(\tau)= {1 \over M} \sum_{k=1}^M
{ {\delta  X_i(t_k+\tau|t_k)} \over  {\delta  X_j(0)}}  \,\, .
$$

Usually, in non Hamiltonian systems, the shape of $\rho({\bf X})$
is not known, therefore relation (\ref{I.4}) does not give a very detailed
information. 
On the other hand the above relation shows that, anyway,  there exists  a connection between
the mean response function $R_{ij}$ and some suitable correlation
function, computed in the unperturbed systems.

In the case of multivariate Gaussian distribution, 
$\ln \rho({\bf X})= -{1 \over 2} \sum_{i,j}\alpha_{ij}X_i X_j + const.$
where $\{ \alpha_{ij} \}$ is a positive symmetric matrix,  
the elements of the linear response matrix can be written in terms
of the usual  correlation functions, $C_{ik}=\langle X_i(t)X_k(0) \rangle / \langle X_i X_k \rangle$, as:
\begin{equation}
\label{I.5}
R_{ij} (t) =\sum_k \alpha_{jk}
{\Bigl \langle X_i(t) X_k(0) \Bigr \rangle }  \; .
\end{equation}
One important nontrivial class of systems with a Gaussian invariant
measure is the inviscid hydrodynamics\footnote{There exist also inviscid 
hydrodynamic systems with non quadratic conservation laws, and, therefore, non 
Gaussian invariant measure. Such cases can have relevance in the statistical mechanics 
of fluids (Pasmanter, 1994).}, 
where the Liouville theorem
holds, and a quadratic invariant exists (Kraichnan, 1959, Kraichnan and Montgomery, 1980,
Bohr et al., 1998). Sometimes in the applications, in absence of detailed information about the shape of $\rho$, 
formula (\ref{I.5}) is assumed to hold to some extent. Numerical studies of simplified models which mimic the chaotic
behavior of turbulent fluids show that, since that stationary
probability distribution is not Gaussian,  Eq. (\ref{I.5})
does not hold. On the other hand, the correlation functions
and the response functions have similar quantitative behavior. 
In particular, in fully developed turbulence, as  one can expect on intuitive ground, 
one has  that the times characterizing the responses  
 approximate the characteristic
correlation times (Biferale et al., 2002,
Boffetta et al., 2003).
This is in agreement with numerical investigation (Kraichnan, 1966)
at moderate Reynolds number of the  Direct Interaction Approximation
equations, showing that, although $R_{ii}(t)$ is not exactly proportional to 
 the autocorrelation function $C_{ii}(t)$, if one compares the correlation times
$\tau_C(k_i)$ (e.g. the time after which the correlation function becomes
lower than  $1/2$)
and the response time $\tau_R(k_i)$ (e.g. the time after which the 
response function becomes lower than  $1/2$),
the ratio $\tau_C(k_a)/\tau_R(k_a)$ remains constant through the inertial
range. In the turbulence context, $X_i$ indicates the  Fourier component of the
velocity field corresponding to a wave vector $k_i$. 

We would like briefly to remark a subtle point. 
From a rather general argument (see Appendix B),
  one has that all the (typical) correlation
functions, at large time delay, have to relax to zero with the same
characteristic time, related to spectral properties of the  operator
$\hat{\bf L}$ which rules the time evolution of the 
probability density function $P({\bf X},t)$:
\begin{equation} 
\label{I.6}
{ \partial \over {\partial t}} P({\bf X},t)=
\hat{\bf L} P({\bf X},t) \, .
\end{equation}
Using  this result in a blind way, 
one has the apparently paradoxical conclusion that, in any kind of 
systems, 
all the correlation functions, relative to degrees of freedom at different scales, 
relax to zero with the same characteristic time. 
On the contrary, in systems with many different characteristic times
(e.g. fully developed turbulence), 
one expects a whole hierarchy of
times distinguishing the behavior at different scales (Frisch, 1995). 
The paradox is, of course, only apparent
since the above  argument is valid just at very long times,
i.e. much longer than the longest characteristic time, and therefore, 
in systems with fluctuations over many different time-scales, this is not very helpful.

\section{Response of fast and slow variables}
\label{sec:detmodel}

Systems with a large number of components and/or with many time scales, e.g. climate dynamics models, 
present clear practical difficulties if one wants to understand their behavior in detail. 
Even using modern supercomputers, it is not possible to simulate all the relevant scales of the  
climate dynamics, which involves processes with characteristic times ranging from days (atmosphere) 
to $10^2$-$10^3$ years (deep ocean and ice shields), see Majda et al. (2005) and Majda and Wang (2006). 

For the sake of simplicity, we consider the 
case in which the state variables evolve over two very different time scales:  
\begin{equation}
\label{eq:slow}
{{d{\bf X}_s} \over {dt}}= {\bf f}({\bf X}_s,{\bf X}_f)
\end{equation}
\begin{equation}
\label{eq:fast}
{{d{\bf X}_f} \over {dt}}= \frac{1}{\epsilon} {\bf g}({\bf X}_s,{\bf X}_f)
\end{equation}
where ${\bf X}_s$ and ${\bf X}_f$ indicate the slow and fast state vectors, respectively,  
$\epsilon \ll 1$ is the ratio between fast and slow characteristic times, 
and both ${\bf f}$ and ${\bf g}$ are $O(1)$. 
A rather general 
issue is to understand the role of the fast variables in the slow dynamics. From the 
practical point of view, one basic question is 
to derive effective equations for the slow variables, e.g. the climatic observable, in which 
the effects of the fast variables, e.g. high frequency forcings, are taken into account by means 
of stochastic parameterization.  
Under rather general conditions (Givon et al., 2004), 
one has the result that, in the limit of small $\epsilon$, the 
slow dynamics is ruled by a Langevin equation with multiplicative noise:
\begin{equation}
\frac{d{\bf X}_s}{dt} = {\bf f}_{eff}({\bf X}_s) + \widehat{\sigma}({\bf X}_s) {\bf \eta}
\label{eq:eff}	
\end{equation}
where ${\bf \eta}$ is a white-noise vector, i.e. its components are Gaussian processes such that 
$\langle \eta_i(t) \rangle=0$ and $\langle \eta_i(t) \eta_j(t') \rangle = \delta_{ij} \delta(t-t')$. 
Although there exist general mathematical results (Givon et al., 2004) 
on the possibility to derive eq. (\ref{eq:eff}) 
from (\ref{eq:slow}) and (\ref{eq:fast}), in practice one has to invoke (rather crude) 
approximations based on physical intuition to determine the shape 
of ${\bf f}_{eff}$ and $\widehat{\sigma}$ (Mazzino et al., 2005). At this regard, see also 
the contribution to the volume by Imkeller and 
von Storch (2001) about stochastic climate models. 
For a more rigorous approach in some climate problems see Majda et al. (1999, 2001) and 
Majda and Franzke (2006). 

In the following, we analyse and discuss two models which, 
in spite of their apparent simplicity, contain the basic features, and the same 
difficulties, of the general multiscale approach:  the Lorenz-96 model (Lorenz 1996) and 
a double-well potential with deterministic chaotic forcing.

\subsection{The Lorenz-96 model}
\label{sec:TheLorenz96Model}

First, let us consider the 
Lorenz-96 system (Lorenz 1996), introduced as a simplified model for the 
atmospheric circulation. Define the set $\{x_k(t)\}$, for 
$k=1,...,N_k$, and  $\{y_{k,j}(t)\}$, for $j=1,...,N_j$, as the slow large-scale variables and 
the fast small-scale variables, respectively (being $N_k=36$ and $N_j=10$). 
Roughly speaking, the $\{x_k\}$'s represent the synoptic scales while the $\{y_{k,j}\}$'s 
represent the convective scales. 
The forced dissipative equations of motion are: 
\begin{equation} 
 \frac{dx_k}{dt} = -x_{k-1}(x_{k-2}-x_{k+1}) - \nu x_k + F + c_1 \sum_{j=1}^{N_j} y_{k,j}
\label{eq:lorenz961}
\end{equation}
\begin{equation} 
\frac{dy_{k,j}}{dt} = -cby_{k,j+1}(y_{k,j+2}-y_{k,j-1}) -c \nu y_{k,j} + c_1 x_k 
\label{eq:lorenz962}
\end{equation}
where: $F=10$ is the forcing term, $\nu=1$ is the linear damping coefficient, $c=10$ is the ratio 
between slow and fast characteristic times, $b=10$ is the relative amplitude between large scale and 
small scale variables, and $c_1=c/b=1$ is the coupling constant that determines the amount of reciprocal 
feedback. 

\begin{figure}[htb]
  \vspace*{2mm}
  \begin{center}
    \includegraphics[width=0.9\textwidth]{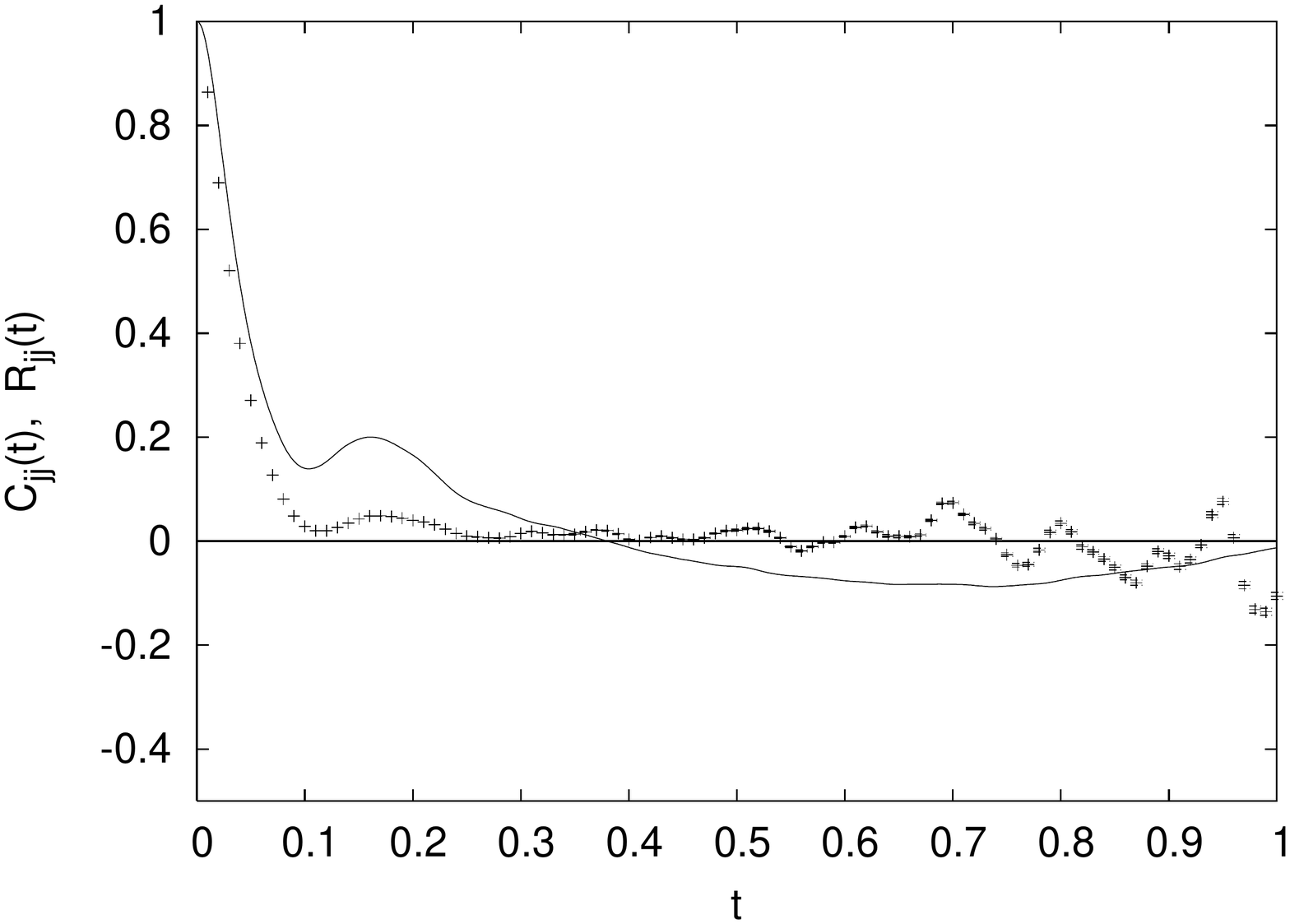}
  \end{center}
  \caption{
  \label{fig:lor96frfast} 
  Lorenz-96 model: autocorrelation $C_{jj}(t)$ (full line) and self-response $R_{jj}(t)$ ($+$)
  of the fast variable $y_{k,j}(t)$ ($k=3,j=3$). The statistical error bars on $R_{jj}(t)$ are of the same 
  size as the graphic symbols used in the plot.
}
\end{figure}

\begin{figure}[htb]
  \vspace*{2mm}
  \begin{center}
    \includegraphics[width=0.9\textwidth]{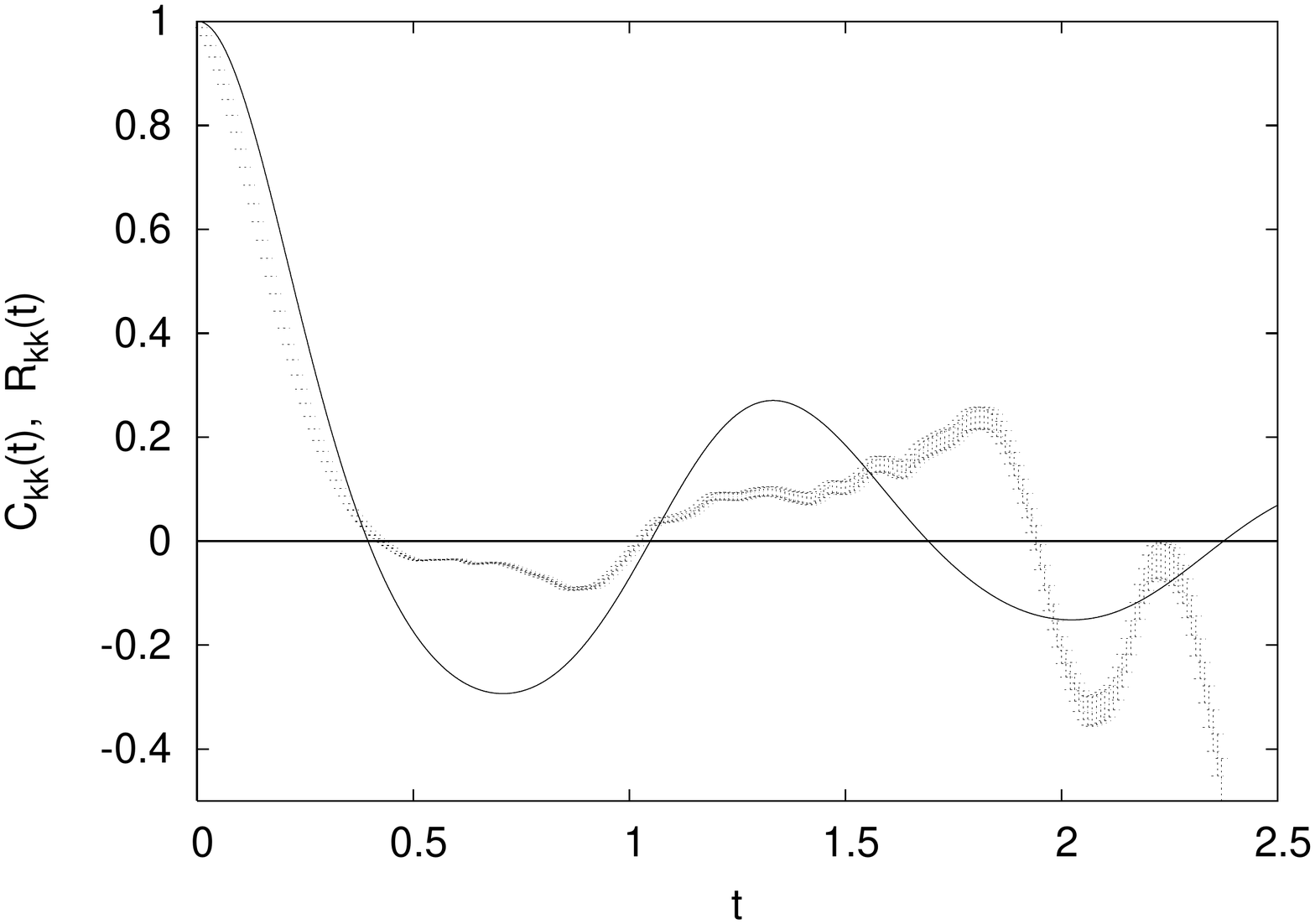}
  \end{center}
  \caption{\label{fig:lor96frslow} 
  Lorenz-96 model: autocorrelation $C_{kk}(t)$ (full line) and self-response $R_{kk}(t)$,  
  with statistical error bars, of the slow variable $x_{k}(t)$ ($k=3$).
}
\end{figure}

Let us consider, first, the response properties of fast and slow variables, 
see Figs. \ref{fig:lor96frfast} and \ref{fig:lor96frslow}. 


In Fig. \ref{fig:lor96frfast}, 
the autocorrelation $C_{jj}(t)$ and self-response $R_{jj}(t)$ refer to the fast variable $y_{k,j}(t)$, 
with fixed $k$ and $j$. It is well evident how, even in absence of a precise agreement between 
autocorrelations and self-response functions (due to the non Gaussian character of the system), 
one has that the correlation of the slow (fast) variables 
have at least a qualitative resemblance with the response of the slow (fast) variables themselves. 

The structure of the Lorenz-96 model includes a rather natural set of quantities  
that suggests how to parameterize the effects of the fast variables on the slow variables, 
for each $k$. Let us indicate with $z_k=\sum_{j=1}^{N_j} y_{k,j}$ the term containing all the 
 $N_j$ fast terms in the equations for the $N_k$ slow modes. In the following, we will see that,  
 replacing the deterministic terms $\{z_k\}$'s in the equations for the $\{x_k\}$'s with suitable 
 stochastic processes, 
 one obtains an effective model able to reproduce the main statistical features of 
 the slow components of the original system. 
 
\begin{figure}[htb]
  \vspace*{2mm}
  \begin{center}
    \includegraphics[width=0.9\textwidth]{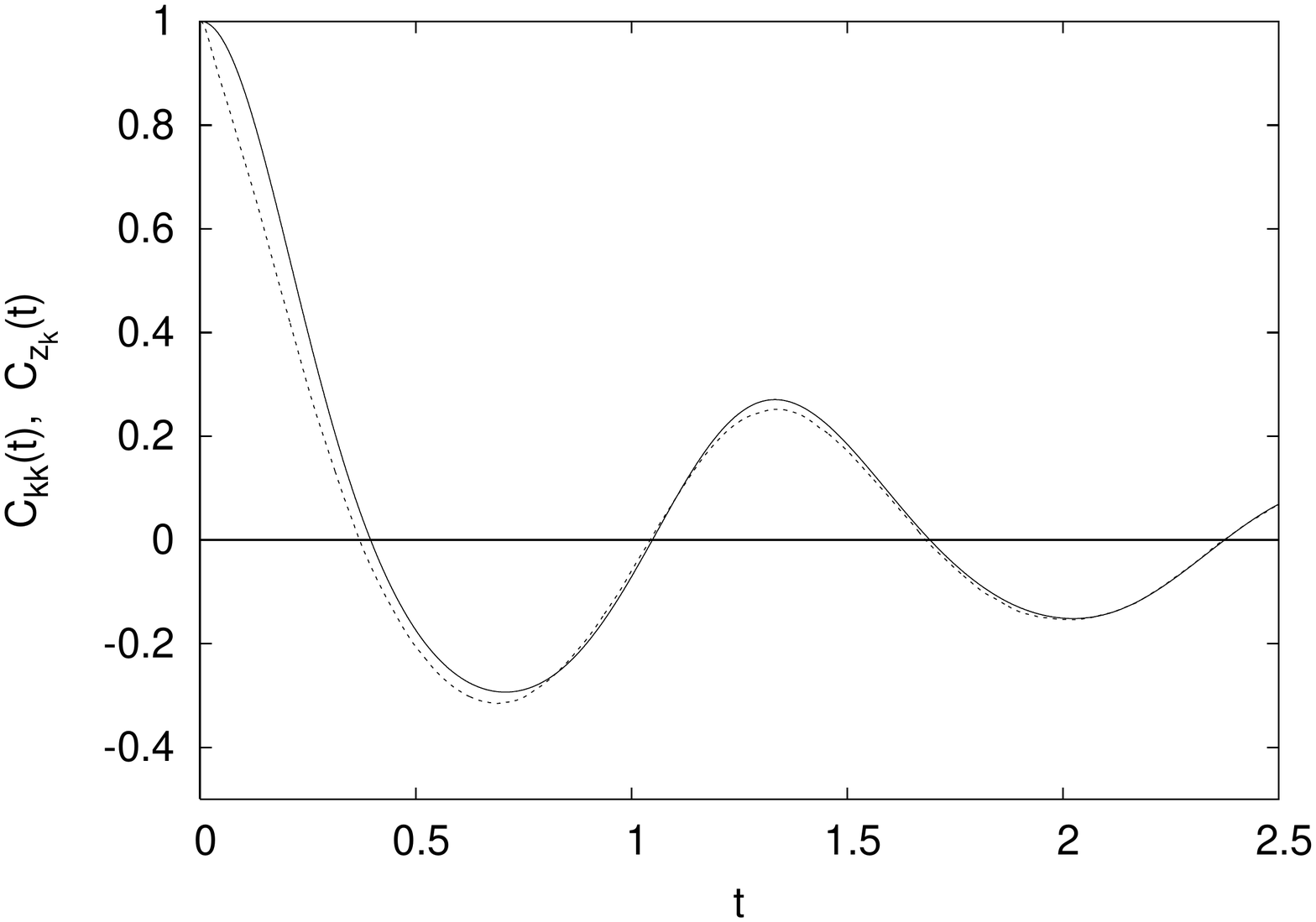}
  \end{center}
  \caption{\label{fig:lor96corrslow} 
  Lorenz-96 model: autocorrelation $C_{z_k}(t)$ (dashed line) of the cumulative variable 
  $z_k(t)$ compared to the autocorrelation $C_{kk}(t)$ of $x_k(t)$ (full line). 
  } 
\end{figure}

It's worth-noting, 
 from Fig. \ref{fig:lor96corrslow}, that $C_{kk}(t)$ and 
$C_{z_k}(t)$, the autocorrelation of the cumulative variable $z_k(t)$, are rather close to each other.  
This suggests that $z_k(t)$ must be  
correlated to $x_k(t)$, in other words, the cumulative effects of the 
$N_j$  fast variables $y_{k,j}(t)$ on $x_k(t)$ are equivalent to an effective slow term, 
proportional to $x_k(t)$. 

We look, therefore, for a conditional white noise parameterization that takes into 
account this important information given by the structure of the Lorenz-96 model equations. 
Let us write the effective equations for the slow modes as
\begin{equation} 
\frac{dx_k}{dt} = -x_{k-1}(x_{k-2}-x_{k+1}) - (\nu + \nu ') x_k + (F+F') + c_2 \cdot \eta_k 
\label{eq:lorX96b}
\end{equation}
where $\eta_k$ are uncorrelated and normalized white-noise terms. 
Some authors, Majda et al (1999, 2001) and Majda and Franzke (2006), using multiscale methods, 
have obtained effective Langevin equations for the slow variables of systems having the same structure 
as the Lorenz-96 model. 
 
\begin{figure}[htb]
  \vspace*{2mm}
  \begin{center}
    \includegraphics[width=0.9\textwidth]{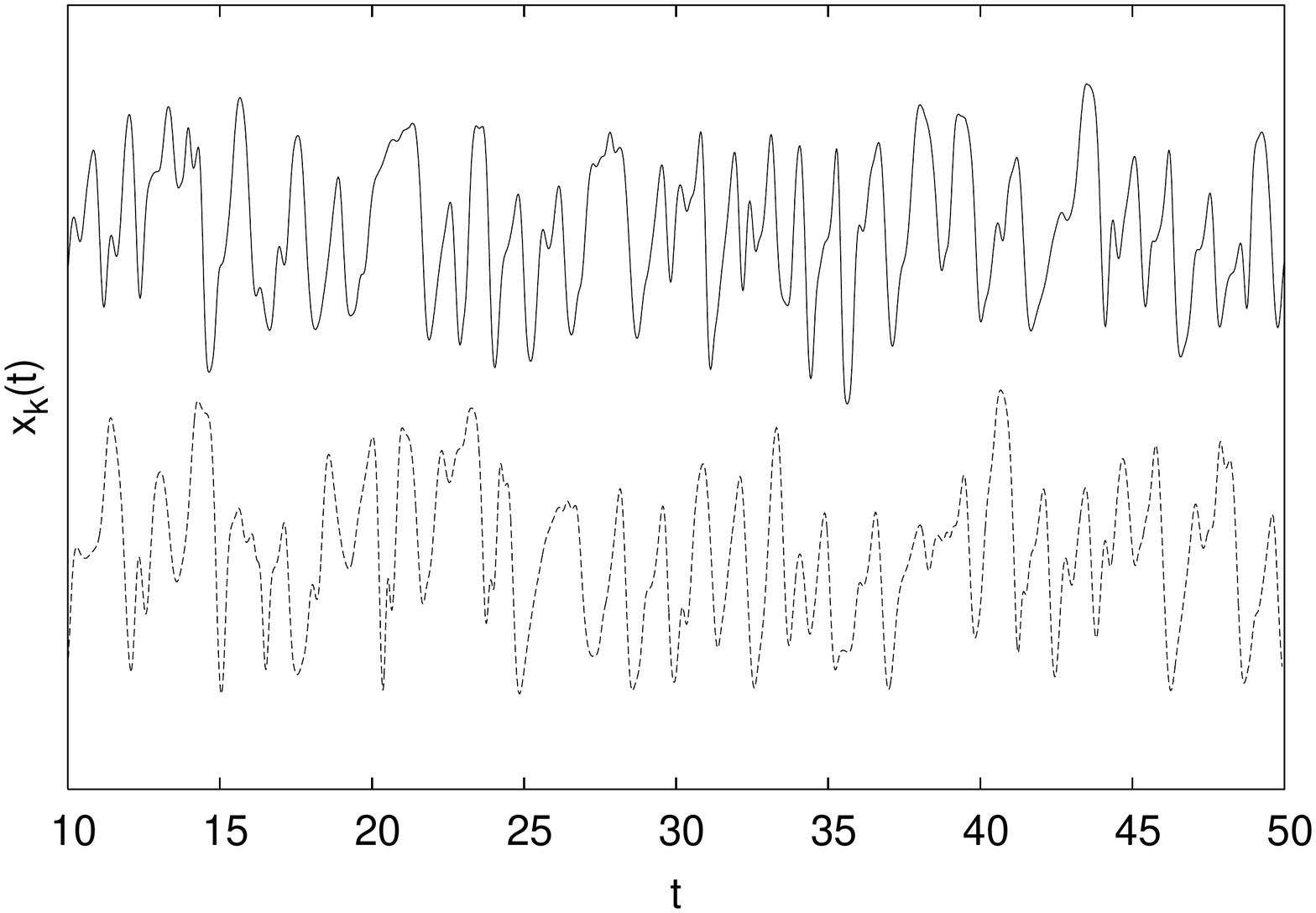}
  \end{center}
  \caption{\label{fig:lor96sigslow} 
  Lorenz-96 model: time signal sample of the slow variable $x_k(t)$ ($k=3$) for the 
  deterministic model (full line) and for the stochastic model (dashed line). For clarity, the two 
  signals have been shifted from each other along the vertical axis.
}
\end{figure}

\begin{figure}[htb]
  \vspace*{2mm}
  \begin{center}
    \includegraphics[width=0.9\textwidth]{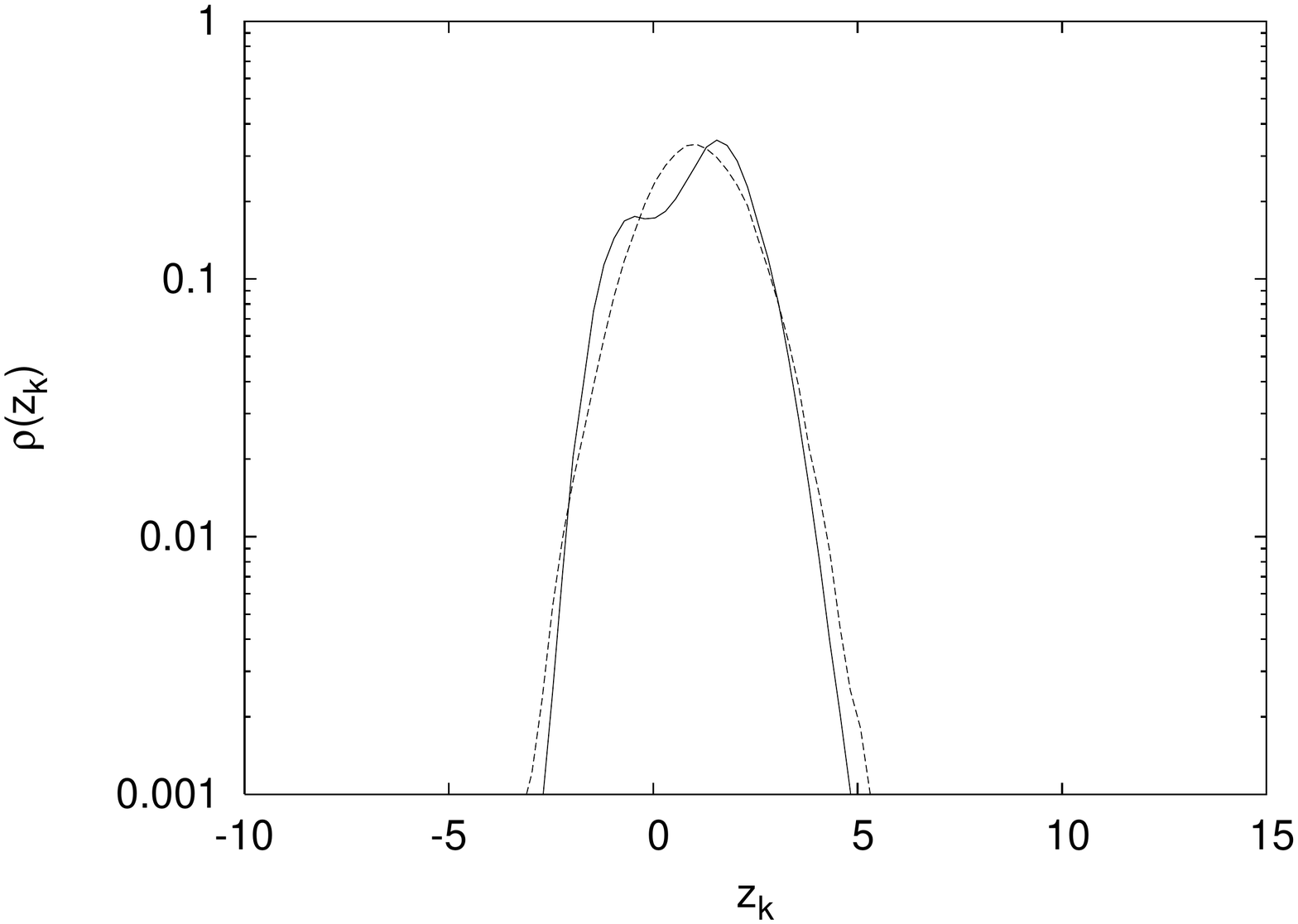}
  \end{center}
  \caption{\label{fig:lor96pdffast} 
  Lorenz-96 model: PDFs of the cumulative variable $z_k$ ($k=3$), 
  see definition in the text for the two cases, for the 
  deterministic model (full line) and the stochastic model (dashed line).
}
\end{figure}

\begin{figure}[htb]
  \vspace*{2mm}
  \begin{center}
    \includegraphics[width=0.9\textwidth]{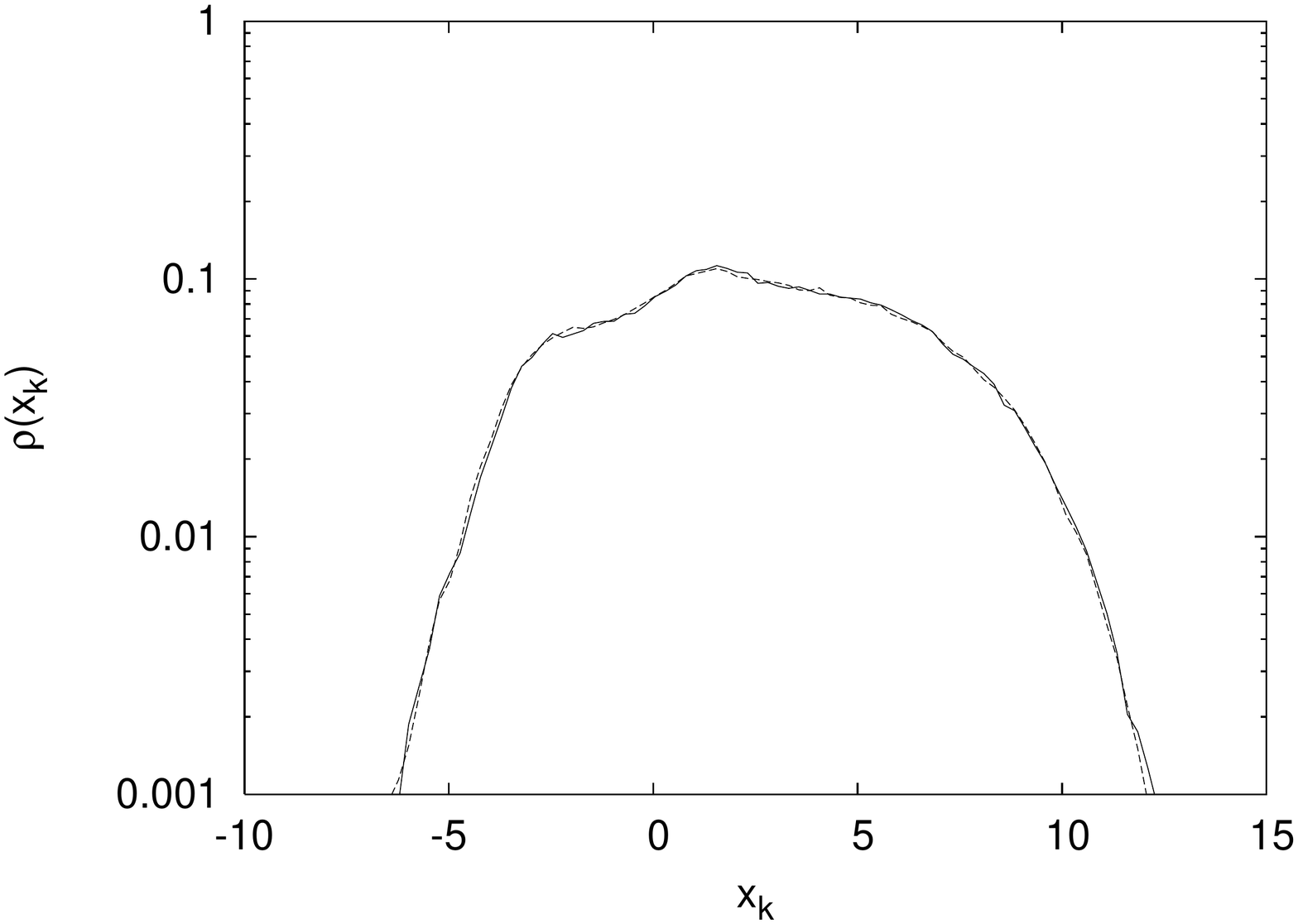}
  \end{center}
  \caption{\label{fig:lor96pdfslow} 
  Lorenz-96 model: PDFs of the slow variable $x_k$ ($k=3$) for the 
  deterministic model (full line) and the stochastic model (dashed line).
}
\end{figure}

Basically we can say that, in the effective model for the slow variables, one parameterizes the effects of the 
fast variables with a suitable renormalization of the forcing, $F \to F+F'$, of the viscosity, 
$\nu \to \nu + \nu'$, and the addition of a random term. 
In other words, we replace the $z_k=\sum_{j=1}^{N_j} y_{k,j}$ terms in (\ref{eq:lorenz961}) with 
 stochastic processes $\widetilde{z_k}$ depending on the slow variables $x_k$:
\begin{equation}
	\frac{dx_k}{dt} = -x_{k-1}(x_{k-2}-x_{k+1}) - \nu x_k + F + \widetilde{c_1} \widetilde{z_k}
	\label{eq:lorX96w}
\end{equation}
where
\begin{equation} 
\widetilde{z_k} = \frac{1}{\widetilde{c_1}} \left( - \nu ' x_k + F' + c_2 \eta_k \right) 
\label{eq:zetak}
\end{equation}
with $\widetilde{c_1}$ is a new coupling constant. We notice that eq. (\ref{eq:lorX96w}) 
has the same form of eq. (\ref{eq:lorenz961}).  
With a proper choice of $\nu '$, $F'$ and $c_2$ in (\ref{eq:lorX96b}), 
$\nu ' = -0.3$, $F'=0.25$, $c_2=0.3$, which implies $\widetilde{c_1}=0.25$ in (\ref{eq:lorX96w}), 
one can reproduce the statistics of $x_k$ and $z_k$ 
to a very good extent, see at this regard 
 Fig. \ref{fig:lor96sigslow}, Fig. \ref{fig:lor96pdffast} and Fig. \ref{fig:lor96pdfslow}. 
 Of course the above described parameterization of the fast variables is inspired to the general 
 `philosophy' of the Large-Eddy Simulation of turbulent geophysical flows at high Reynolds numbers 
 (Moeng, 1984, Moeng and Sullivan, 1994, Sullivan et al., 1994).



\begin{figure}[htb]
\vspace*{2mm}
\begin{center}
 \includegraphics[width=0.9\textwidth]{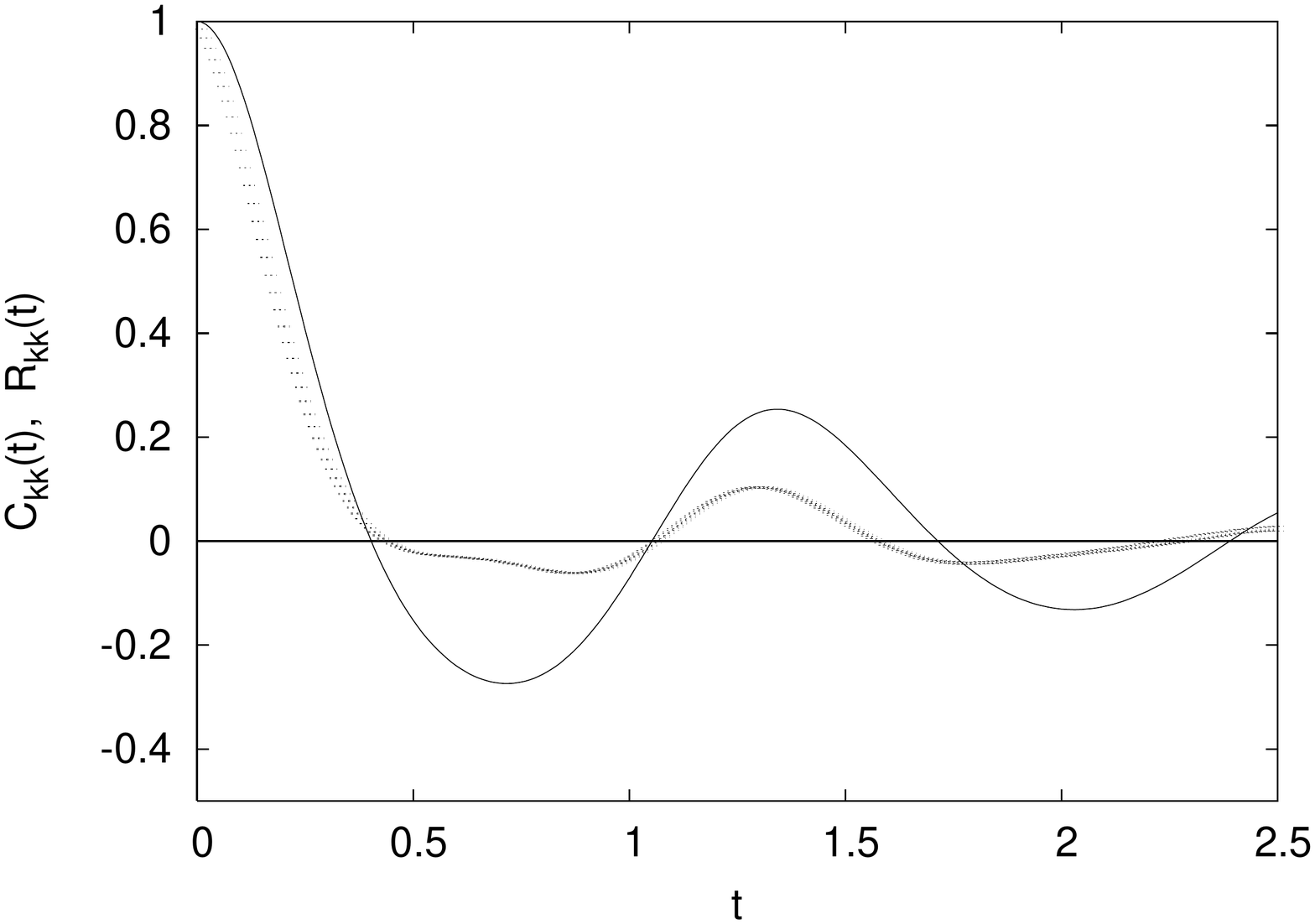}
\end{center}
\caption{\label{fig:lor96frnoise} 
Lorenz-96 model: autocorrelation $C_{kk}(t)$ (full line) and self-response 
 $R_{kk}(t)$, with statistical error bars, of the slow variable $x_k(t)$ for the stochastic model. 
 }
\end{figure}

The FR properties of the stochastic Lorenz-96 slow variables are reported in 
Fig. \ref{fig:lor96frnoise}.

\begin{figure}[htb]
  \vspace*{2mm}
  \begin{center}
    \includegraphics[width=0.9\textwidth]{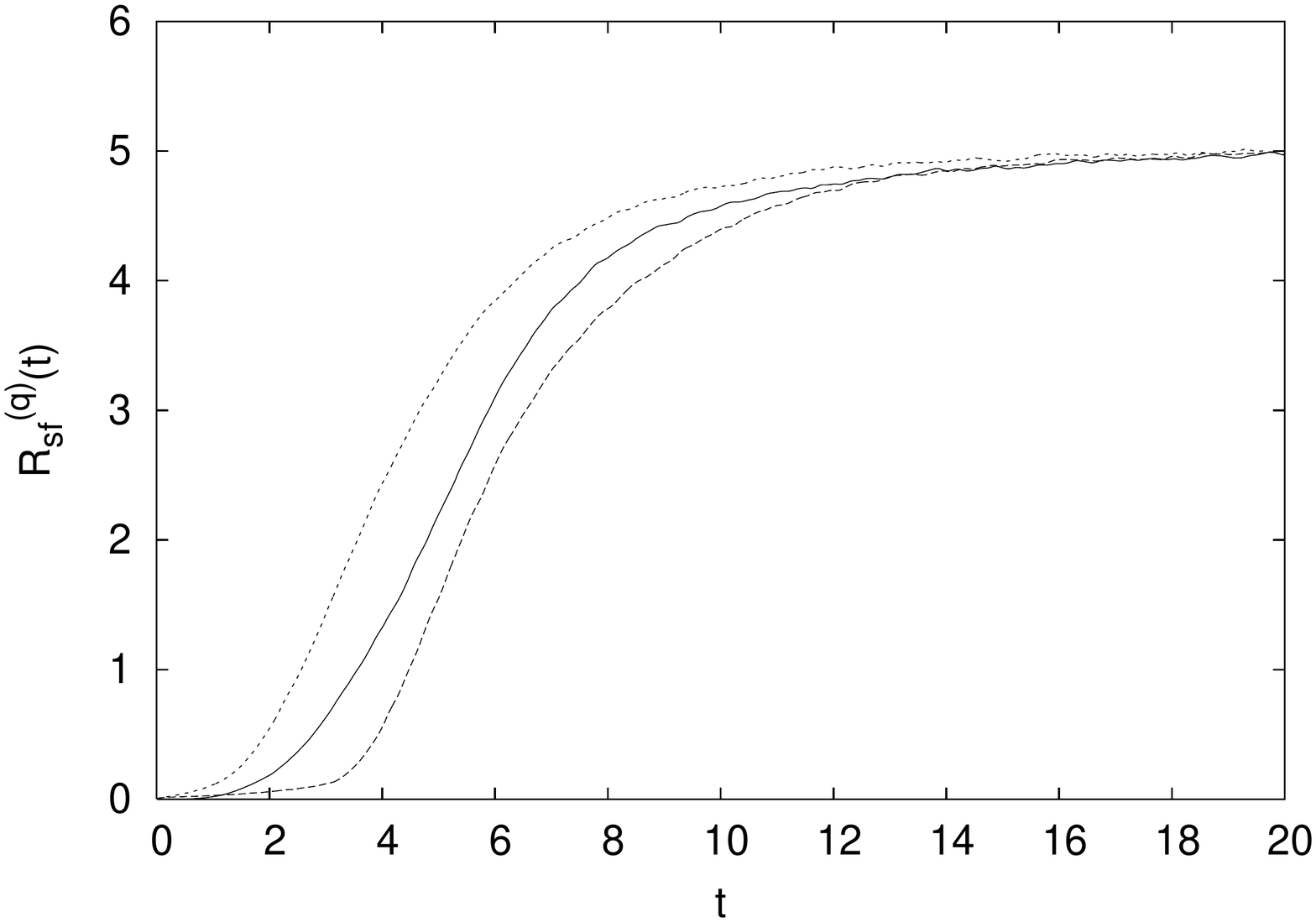}
  \end{center}
  \caption{\label{fig:lor96crfslow} 
Lorenz-96 model: quadratic cross-response function $R_{sf}^{(q)}(t)$ 
  for the deterministic model (full line), for the stochastic model when 
  the slow variables 
  evolve with the same noise realization for all components except one (dashed line), 
  and when the slow variables evolve with a different noise realization 
  for every component (dotted line).
  }
\end{figure}

Let us come back to the response problem. Of course the 
mean response of a slow variable to a perturbation on a fast variable is zero. However, this does not mean that 
the effect of the fast variables on the slow dynamics is not statistically relevant. Let us 
introduce the quadratic response of $x_k(t)$ with respect to an $infinitesimal$ perturbation on 
$y_{k,j}(0)$, for fixed $k$ and $j$: 
\begin{equation}
R^{(q)}_{kj}(t) = \frac{ \left[\overline{ \delta x_k (t)^2 }\right]^{1/2}}{\delta y_{k,j}(0)}	
\label{eq:rqj}
\end{equation}
Considered that in all simulations the initial impulsive perturbations on the $y_{k,j}$ is kept 
constant, $\delta y_{k,j}(0) = \Delta$, with $\Delta \ll \langle y_{k,j}^2 \rangle^{1/2}$, 
it is convenient to  
take the average of (\ref{eq:rqj}) over all $j$'s, at a fixed $k$, and introduce the quantity:  
\begin{equation}
R^{(q)}_{sf}(t) = \frac{\Delta}{N_j} \sum_{j=1}^{N_j} R^{(q)}_{kj}(t)
\label{eq:rqm}	
\end{equation}
where with $s$ and $f$ we label the slow and fast variables, respectively. 
In the case 
of the Lorenz-96 system, all the $y_{k,j}$ variables, at fixed $k$, are statistically equivalent, and 
have identical coupling with $x_k$, so that $R^{(q)}_{sf}(t)/\Delta$ coincides with $R^{(q)}_{kj}(t)$.   
 We report in Fig. \ref{fig:lor96crfslow} the behavior of 
$R^{(q)}_{sf}(t)$, for both (\ref{eq:lorenz961}) and (\ref{eq:lorX96w}). 
As regards to the stochastic model, the analogous of (\ref{eq:rqm}) is defined as follows.  
 One studies the evolution of $\delta x_k(t)$ as difference of two trajectories obtained 
 with two different realizations of the $\left\{ \eta_k \right\}$'s. 
 It is worth stressing that the behavior of $\delta x_k(t)$ under two noise realizations can be 
 very different from the behavior of $\delta x_k(t)$ under the same noise realization (see Appendix C). 
 This aspect will be considered again in the next section.

\subsection{A simplified model}
\label{sec:TheDoubleWellModel}

In order to grasp the essence of systems with fast and slow variables, we discuss now 
a toy climate model in which the `climatic' variable fluctuates between two states. 
Consider a four dimensional state vector ${\bf q}=(q_0,q_1,q_2,q_3)$ whose  
evolution is given by:
\begin{equation} 
{dq_0 \over dt} = 2\sqrt{H} q_0 - q_0^3 + cq_1 
\label{eq:dwchaos1}
\end{equation}
\begin{equation}
{dq_1 \over dt} = \frac{1}{\widetilde{\epsilon}} [-\sigma_L (q_1-q_2)] 
\label{eq:dwchaos2}
\end{equation}
\begin{equation}
{dq_2 \over dt} = \frac{1}{\widetilde{\epsilon}} [-q_1q_3+r_Lq_1-q_2]
\label{eq:dwchaos3}
\end{equation}
\begin{equation}
{dq_3 \over dt} = \frac{1}{\widetilde{\epsilon}} [q_1q_2-b_Lq_3] 
\label{eq:dwchaos4}
\end{equation}
This four equation system will be named the deterministic DW model. 
The subsystem formed by  (\ref{eq:dwchaos2}), (\ref{eq:dwchaos3}) and (\ref{eq:dwchaos4}) is nothing but the 
well-known Lorenz-63 model (Lorenz 1963), in which the constant $\widetilde{\epsilon}$  
has the function of rescaling the characteristic time. 
In absence of coupling ($c=0$) between $q_0$ and $q_1$,  
the unforced motion equation holds for the slow variable $x=q_0$: 
\begin{equation}
{dx \over dt} = - \frac{\partial V}{\partial x} 
	= 2\sqrt{H} x - x^3  \ \ \ \ \textnormal{with} \ \ \ \ V(x)=H - \sqrt{H} x^2 + \frac{1}{4} x^4 
	\label{eq:dw}
\end{equation}
The system (\ref{eq:dw}) has one unstable steady state in $x_0=0$ corresponding to the top of the 
hill of height $H$, 
and two stable steady states in $x_{1/2}=\pm (4H)^{1/4}$, i.e. the bottom of the valleys. 
The presence of the coupling ($c \neq 0$) between slow and fast variables 
can induce transitions between the two valleys. 
The parameters in (\ref{eq:dwchaos1}), (\ref{eq:dwchaos2}), (\ref{eq:dwchaos3}), and (\ref{eq:dwchaos4}) 
are fixed to the following values: 
$\sigma_L=10$, $r_L=28$, $b_L=8/3$, i.e. the classical set-up corresponding to the chaotic regime for 
the Lorenz-63 system; 
$H=4$, the height of the barrier; $c=0.5$, the coupling constant 
that rules the transition time scale of $q_0(t)$ between the two valleys; by setting $\widetilde{\epsilon}=1$, 
the ratio $\epsilon$ between fast and slow characteristic times, see (\ref{eq:slow}) and (\ref{eq:fast}), is 
$ O(10^{-1})$.  

\begin{figure}[htb]
  \vspace*{2mm}
  \begin{center}
    \includegraphics[width=0.9\textwidth]{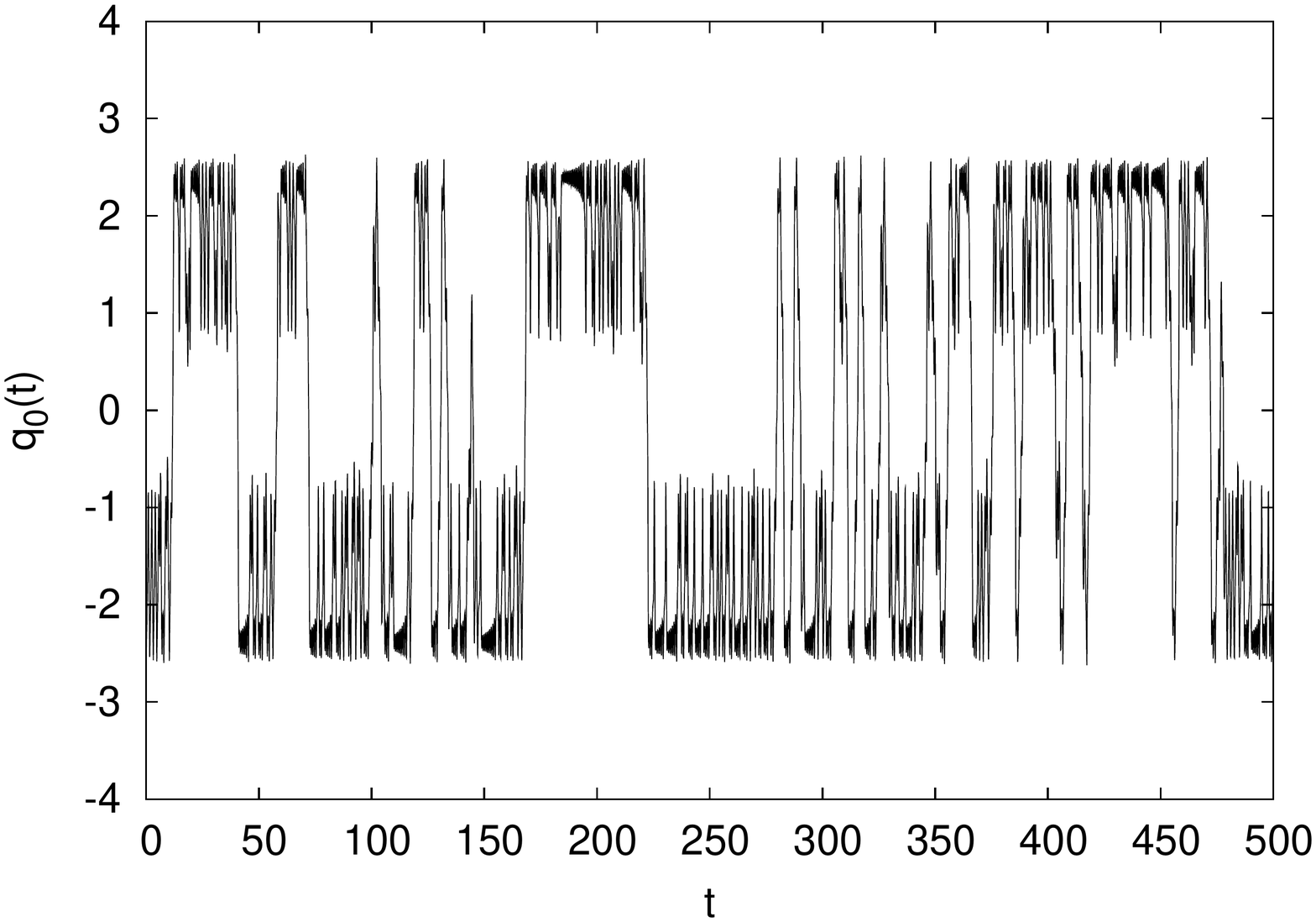}
  \end{center}
  \caption{\label{fig:dwdetq0} 
  DW model with $\widetilde{\epsilon}=1$: 
  time signal sample of the slow variable $q_0(t)$. The ratio between fast and slow characteristic times 
  is $\epsilon \sim 0.1$ (see text).
}
\end{figure}
\begin{figure}[htb]
  \vspace*{2mm}
  \begin{center}
    \includegraphics[width=0.9\textwidth]{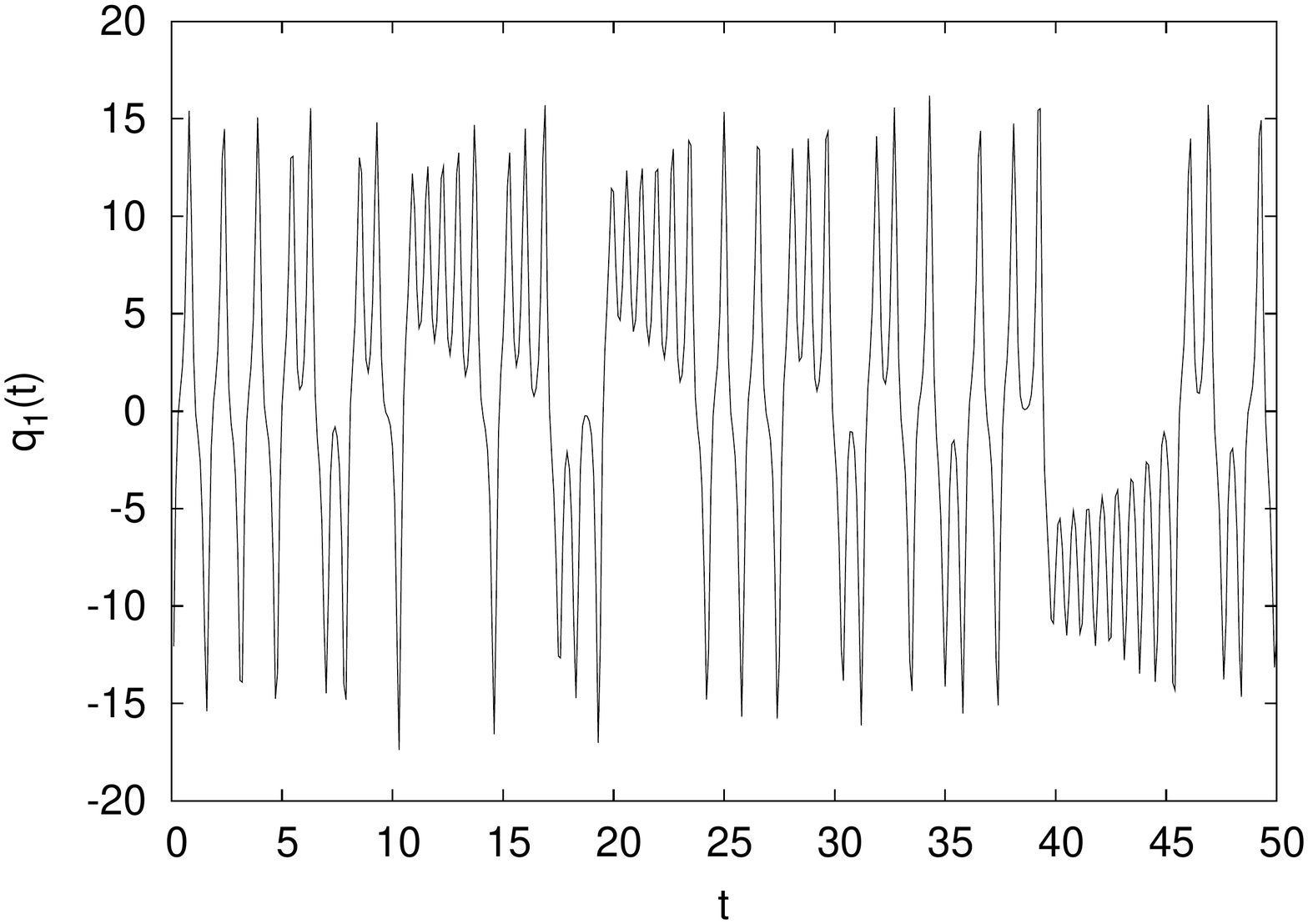}
  \end{center}
  \caption{\label{fig:dwdetq1} 
DW model with $\widetilde{\epsilon} = 1$: time signal sample of the fast variable $q_1(t)$.
}
\end{figure}

Since the time scale of the $q_0(t)$ well-to-well transitions 
may be considerably longer, depending on $1/c$, than the characteristic time of $q_1(t)$, of order $O(1)$, 
we refer to $q_0$ as the slow variable, or the low-frequency observable, and 
to $q_1$ as the fast variable, or the high-frequency forcing, of the deterministic DW model.  
It can be easily shown that, for $c = 0$, small  
perturbations $\Delta q_0$ around the two potential minima at $\pm (4H)^{1/4}$ 
relax exponentially to zero with characteristic time $1/4\sqrt{H}$.  
For sufficiently large values of $c$, the climatic variable $q_0(t)$ 
jumps aperiodically back and forth between the two valleys, driven by the chaotic signal $q_1(t)$, see 
Fig. \ref{fig:dwdetq0} and Fig. \ref{fig:dwdetq1}. 


The main statistical quantities investigated to analyse the DW model are the following:  

a) the probability density function of the slow variable $q_0$;

b) the probability density function of the well-to-well transition time $t_e$, $\rho (t_e)$;
 
c) the slow and fast auto-correlation functions (ACF) 
$C_{ii}(t)=\langle q_i(t)q_i(0) \rangle / \langle q_i^2 \rangle$, with $i=0,1$; 

d) the slow and fast self-response functions (ARF) 
 $R_{ii}(t)= \overline{\delta q_i(t) / \delta q_i(0)} $, with $i=0,1$; 
  
e) the quadratic cross-response function of the slow variable  
 $q_0(t)$ with respect to the fast variable $q_1(0)$.

Of course $R_{01}(t)$, i.e. the mean response of $q_0(t)$ to a perturbation on $q_1(0)$, is zero for 
trivial symmetry arguments. On the other hand, the quadratic response:
\begin{equation}
R^{(q)}_{01}(t) = { \left[\overline{ \delta q_0(t)^2 } \right]^{1/2} \over \delta q_1(0)}
\label{eq:quadR}	
\end{equation}
can give relevant physical information. Even in this case, since in all simulations the initial 
perturbation on $q_1(0)$ is kept constant, $\delta q_1(0) = \Delta \ll \langle q_1^2 \rangle^{1/2}$, 
it is convenient to define as mean quadratic response of the slow variable ($s$) with respect to the fast 
variable ($f$) the quantity $R^{(q)}_{sf}(t) = \Delta \cdot R^{(q)}_{01}(t)$. The long-time 
saturation level of $R^{(q)}_{sf}(t)$ is of the order of the distance between the two climatic states. 

With the current set-up, slow and fast variable have characteristic times which differ by an order of magnitude 
from each other, while the statistics of $q_0$ is strongly non Gaussian. Because of the skew 
structure of the system, i.e. the fast dynamics drives the slow dynamics but without counter-feedback, 
one expects that, at the least in the limit of large time scale separation, 
the joint PDF can be factorized, with an asymptotic PDF for $q_0$ of the form 
$\rho_{0} = K \cdot e^{-V_{eff}(q_0)}$, where $K$ is a normalization constant. 

\begin{figure}[htb]
  \vspace*{2mm}
  \begin{center}
    \includegraphics[width=0.9\textwidth]{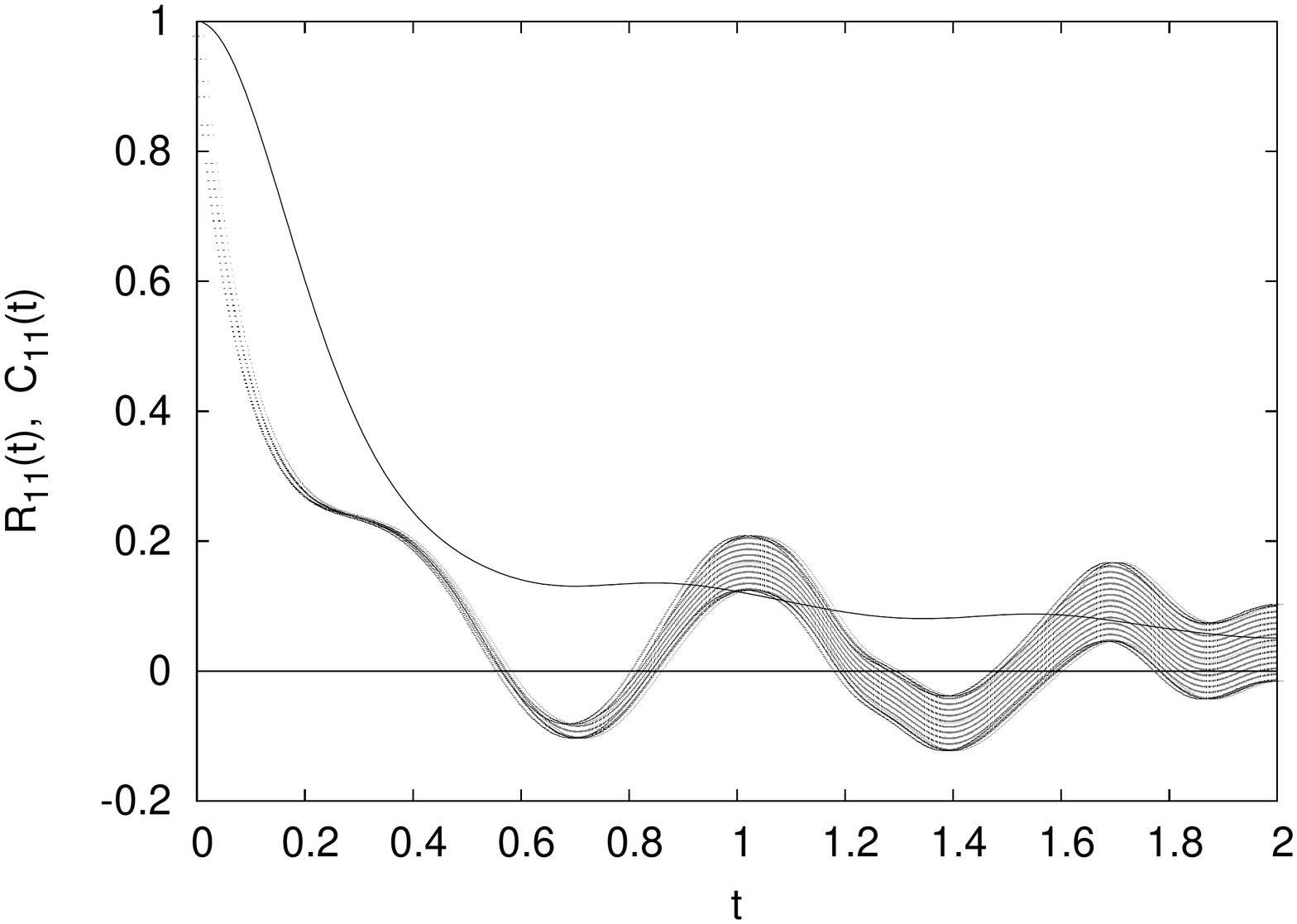}
  \end{center}
  \caption{\label{fig:dwdetfrq1} 
  DW model with $\widetilde{\epsilon} = 1$: autocorrelation $C_{11}(t)$ (full line) 
  and self-response $R_{11}(t)$, with statistical error bars, for the fast variable $q_1$.
}
\end{figure}

\begin{figure}[htb]
  \vspace*{2mm}
  \begin{center}
    \includegraphics[width=0.9\textwidth]{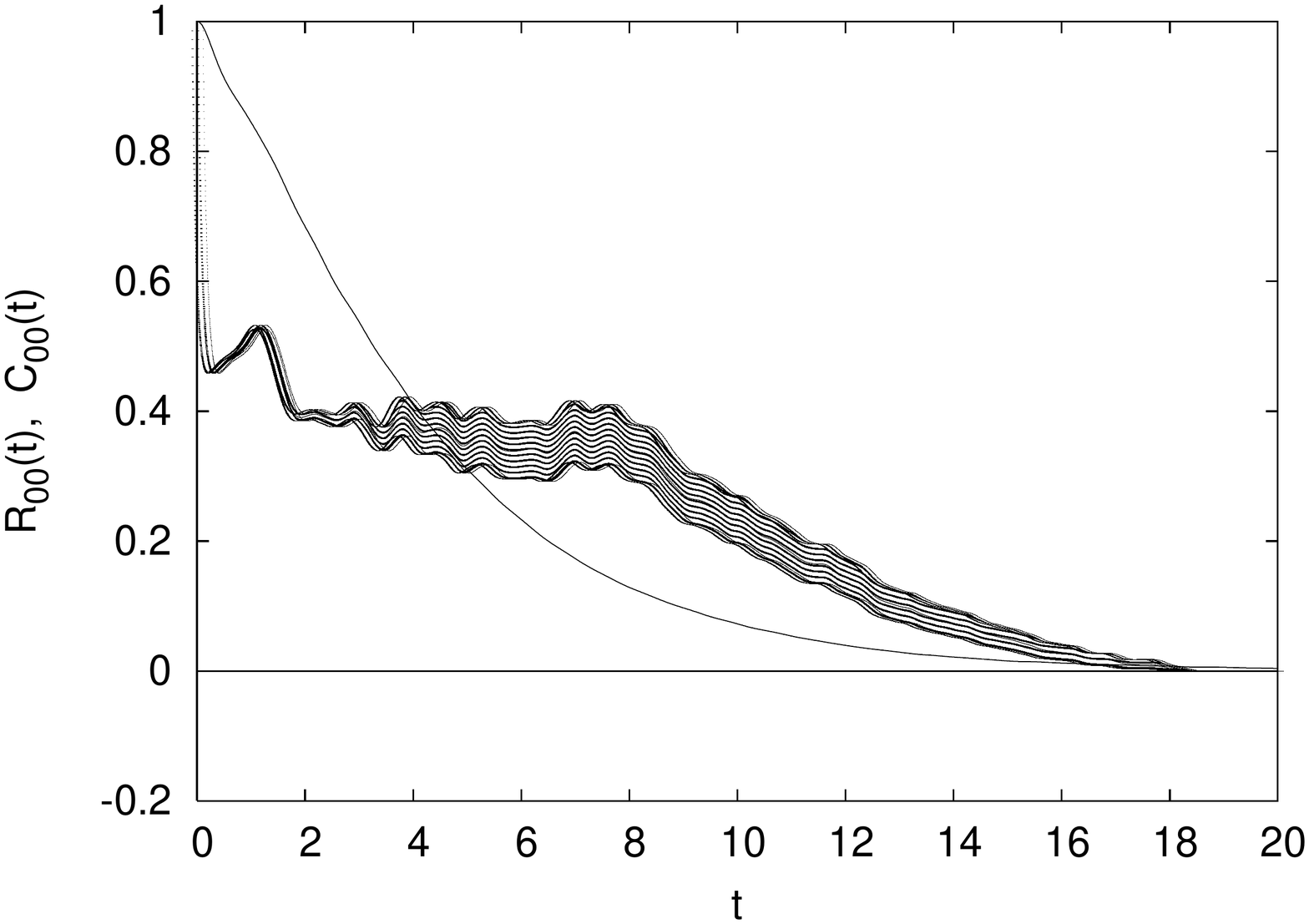}
  \end{center}
  \caption{\label{fig:dwdetfrq0} 
DW model with $\widetilde{\epsilon} = 1$: Autocorrelation $C_{00}(t)$ (full line) and 
  self-response $R_{00}(t)$, with statistical error bars, for the slow variable $q_0$. 
  }
\end{figure}

The FR properties of the deterministic DW model, for the fast and slow variables, 
 are shown in Fig. \ref{fig:dwdetfrq1} and Fig. \ref{fig:dwdetfrq0}, respectively. 
  

The slow self-response $R_{00}(t)$ initially decreases  
exponentially with characteristic time $1/4\sqrt{H}$ ($H=4$), i.e. the same behavior of the  
relaxation of a small perturbation near the bottom of a valley for $c=0$.  
Then,  $R_{00}(t)$ relaxes to zero much more slowly. It is natural to assume that this is due   
 to the long-time jumps between the valleys. It is well evident 
 that $R_{00}$ behaves rather differently from $C_{00}$, while $R_{11}$ and $C_{11}$ have, at least, the 
 same qualitative shape. On the other hand, the autocorrelation (self-response) time scales of the two variables 
 differ from each other of a factor $\sim 10$, compatibly with the fact that 
 the ratio between fast and slow characteristic times is $\epsilon \sim 0.1$,   
 for the current set-up ($\widetilde{\epsilon}=1$). 
 
\begin{figure}[htb]
  \vspace*{2mm}
  \begin{center}
    \includegraphics[width=0.9\textwidth]{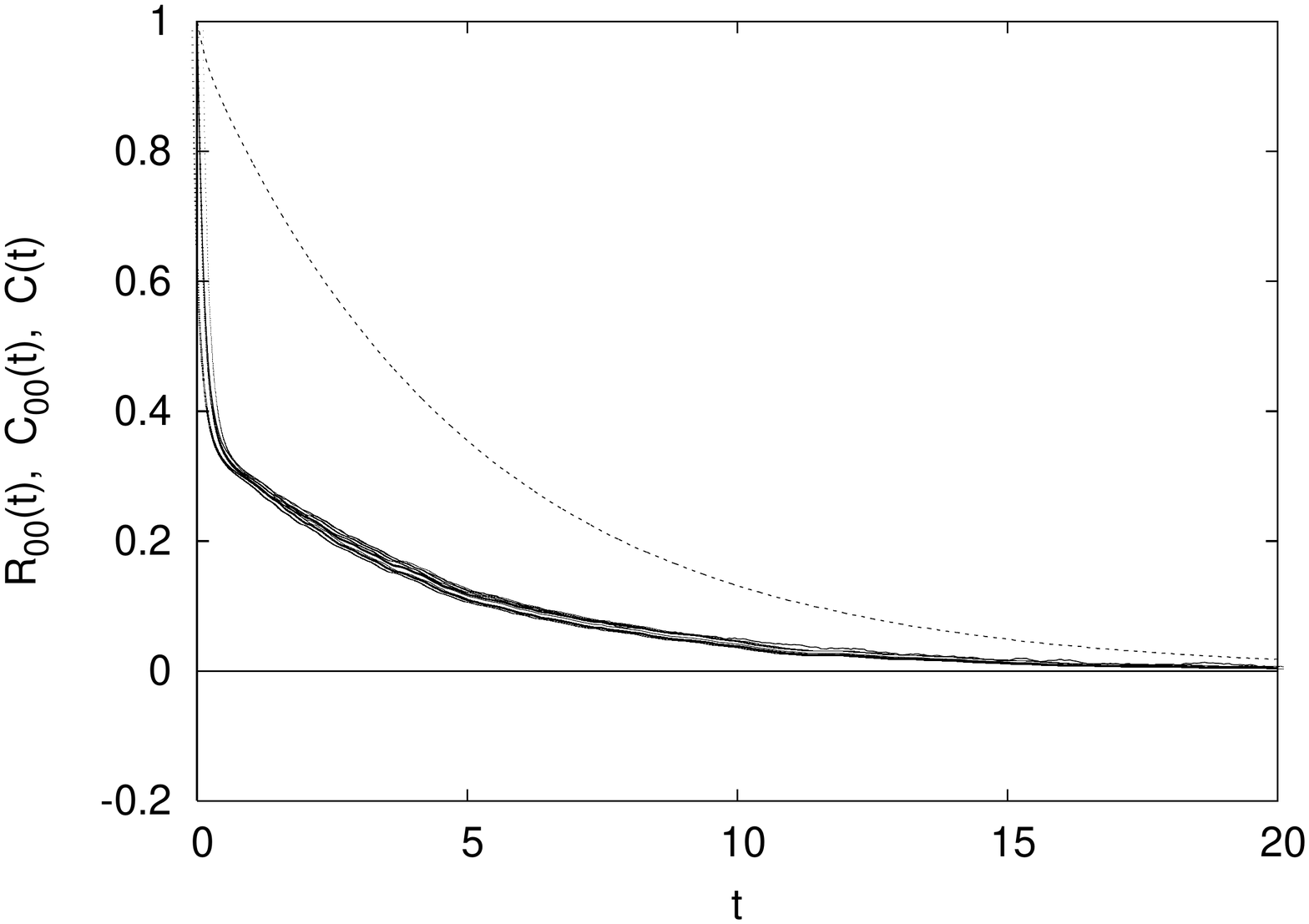}
  \end{center}
  \caption{\label{fig:dwfr1000} 
DW model with $\widetilde{\epsilon} = 0.01$, implying $\epsilon \sim 10^{-3}$: 
  autocorrelation $C_{00}(t)$ (dashed line), self-response $R_{00}(t)$, 
  with statistical error bars, and the correlation function 
  $C(t)$ predicted by the FRR (full line) which is actually undistinguishable from the response.
}
\end{figure}

 Since the statistics is far from being Gaussian, the `correct' correlation function which satisfies the FR theorem, 
 for the slow variable, has the form:
 \begin{equation}
C(t) = - \biggl \langle q_0(t) \left. {\partial \rho_{\epsilon}(q_0,q_1,q_2,q_3) \over \partial q_0 } \right|_{t=0}  \biggr \rangle 
\label{eq:Ctilde} 
\end{equation}
where $\rho_{\epsilon}(q_0,q_1,q_2,q_3)$ is the (unknown) joint PDF of the state variable of the system at 
a fixed $\epsilon$. 
In the limit of large time separation, i.e. for $\widetilde{\epsilon} \to 0$, one expects 
that the asymptotic PDF $\rho_{0}(q_0,q_1,q_2,q_3)$ is factorized:
\begin{equation}
\rho_{0}(q_0,q_1,q_2,q_3)= K e^{-V_{eff}(q_0)} \rho_{L}(q_1,q_2,q_3)
\label{eq:Plimit}	
\end{equation}
where $K$ is a normalization constant, and $\rho_L$ is the PDF of the Lorenz-63 state variable. 
Under this condition, the right correlation function predicted by the FRR has a relatively simple form:
\begin{equation}
C(t) = \biggl \langle q_0(t) \left. 
{\partial V_{eff}(q_0) \over \partial q_0 } \right|_{t=0} \biggr \rangle 
\label{eq:Ctilde2}	
\end{equation}
where $V_{eff}$ indicates the effective potential. 
For $\epsilon \sim 10^{-1}$ (corresponding to $\widetilde{\epsilon}=1$) 
we have checked numerically that the joint PDF is not yet factorized, while 
for a very small ratio between the characteristic times, $\epsilon \sim 10^{-3}$ 
(corresponding to $\widetilde{\epsilon} = 10^{-2}$), 
the form  (\ref{eq:Plimit})	holds and, 
taking $V_{eff} \propto V$, we obtain a very good agreement between $R_{00}(t)$ and $C(t)$, see 
Fig. \ref{fig:dwfr1000}. 

\begin{figure}[htb]
  \vspace*{2mm}
  \begin{center}
    \includegraphics[width=0.9\textwidth]{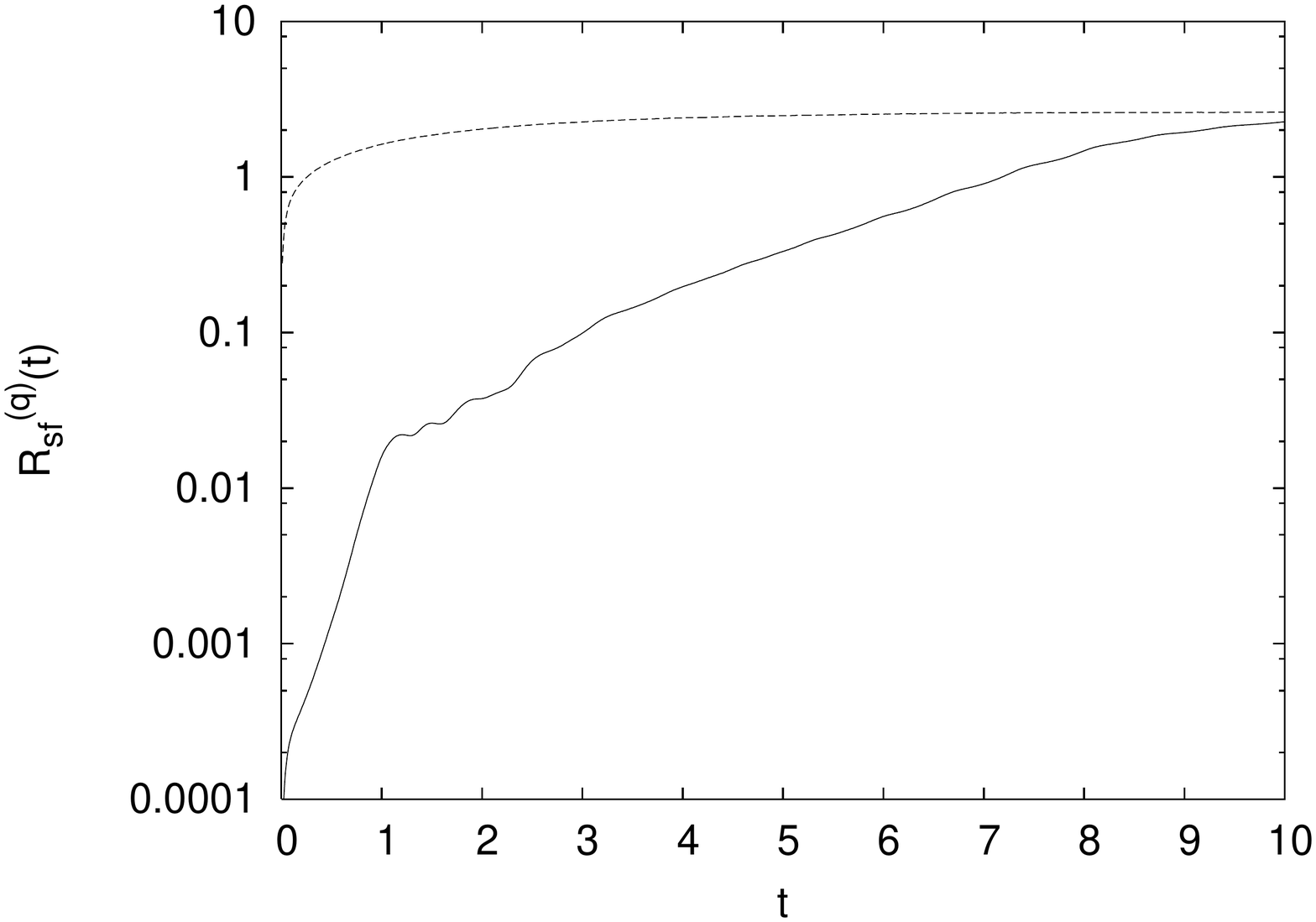}
  \end{center}
  \caption{\label{fig:dwallCRF} 
Quadratic cross-response function $R_{sf}^{(q)}(t)$ for the 
  DW model (full line) and the WNDW model (dashed line). 
  The growth rates of $R_{sf}^{(q)}(t)$ for the DW model   
  are compatible with the two characteristic times of the system, while for the WNDW model 
  $R_{sf}^{(q)}(t)$ quickly saturates in a very short time.
}
\end{figure}

The cross-response properties of the DW model, measured by the quantity $R_{sf}^{(q)}(t)$, are reported 
 in Fig. \ref{fig:dwallCRF}. We will consider again later this issue when discussing the stochastic modeling. 
 While the mean (slow-to-fast) cross-response is null (not shown), its fluctuations grow with time. 
 This means that an initial uncertainty on the fast variables has consequences for the predictability 
 of the slow variable, since it induces a mean separation growth between two initially close `climatic' states 
 of the $q_0$ variable. 
 At small times, $R_{sf}^{(q)}(t)$ grows exponentially in time, 
 i.e. it is driven by the chaotic character of the fast variable while, at 
 very long times, the well-to-well aperiodic jumps play the dominant role and the growth speed 
 eventually decreases to zero until saturation sets in. 
 
\begin{figure}[htb]
  \vspace*{2mm}
  \begin{center}
    \includegraphics[width=0.9\textwidth]{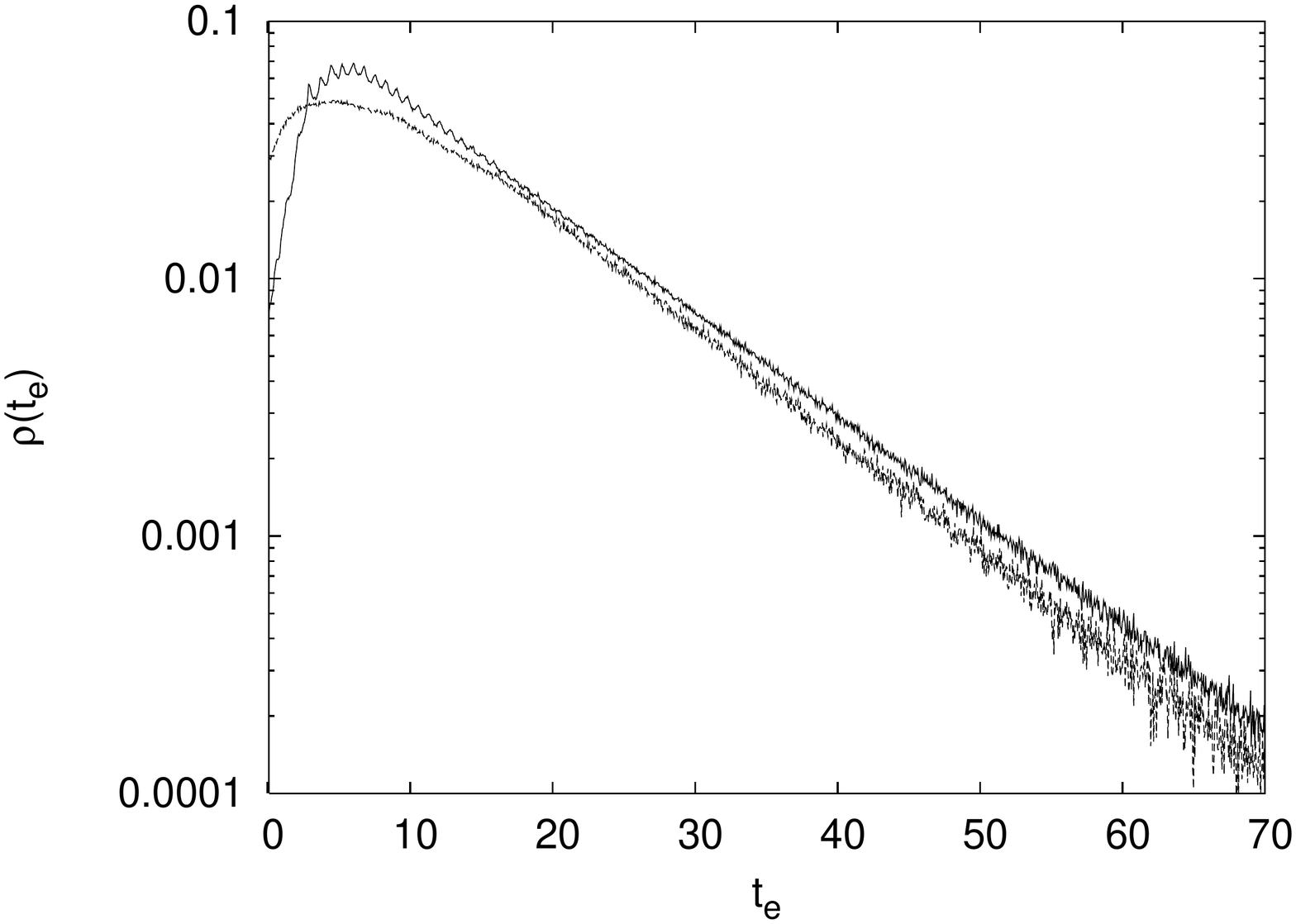}
  \end{center}
  \caption{\label{fig:dwtkpdf} 
Comparison of the PDFs of the transition time $t_e$ between the two climatic states for 
  the DW model (full line) and the WNDW model (dashed line), for $\widetilde{\epsilon} = 1$ ($\epsilon \sim 0.1$).
}
\end{figure}

 Let us now consider a stochastic model for the slow variable $q_0(t)$, 
obtained by replacing the fast variable $q_1$, in the equation for $q_0$, with a 
white noise.   
One has a Langevin equation of the kind: 
\begin{equation} 
{dq_0 \over dt}(t) = 2\sqrt{H}q_0(t)  - q_0^3 + \sigma \cdot \xi(t)
\label{eq:dwnoise1}
\end{equation}
where $\xi(t)$ is a Gaussian process with $\langle \xi(t) \rangle=0$ and 
$\langle \xi(t)\xi(t') \rangle = \delta(t-t')$.  
We call eq. (\ref{eq:dwnoise1}) the WNDW model. The value  
$\sigma=19.75$ is determined by requiring that the PDFs of the well-to-well transition times have the same 
asymptotic behavior (i.e. exponential tail with the same exponent), see Fig. \ref{fig:dwtkpdf}.  

\begin{figure}[ht]
  \vspace*{2mm}
  \begin{center}
    \includegraphics[width=0.9\textwidth]{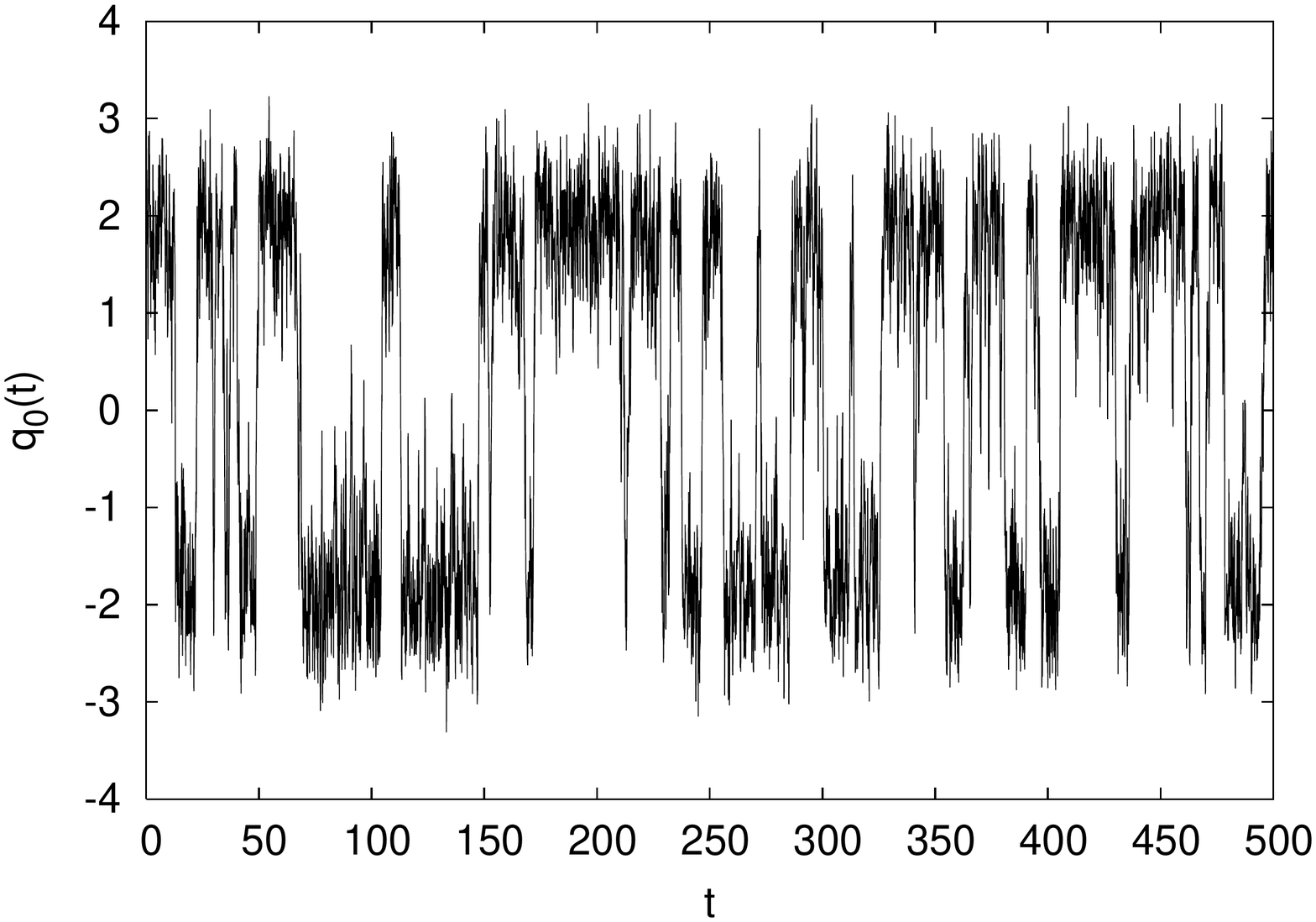}
  \end{center}
  \caption{\label{fig:dwwnoiseq0} 
WNDW model: time signal sample of the slow variable $q_0(t)$.
}
\end{figure}
\begin{figure}[htb]
  \vspace*{2mm}
  \begin{center}
    \includegraphics[width=0.9\textwidth]{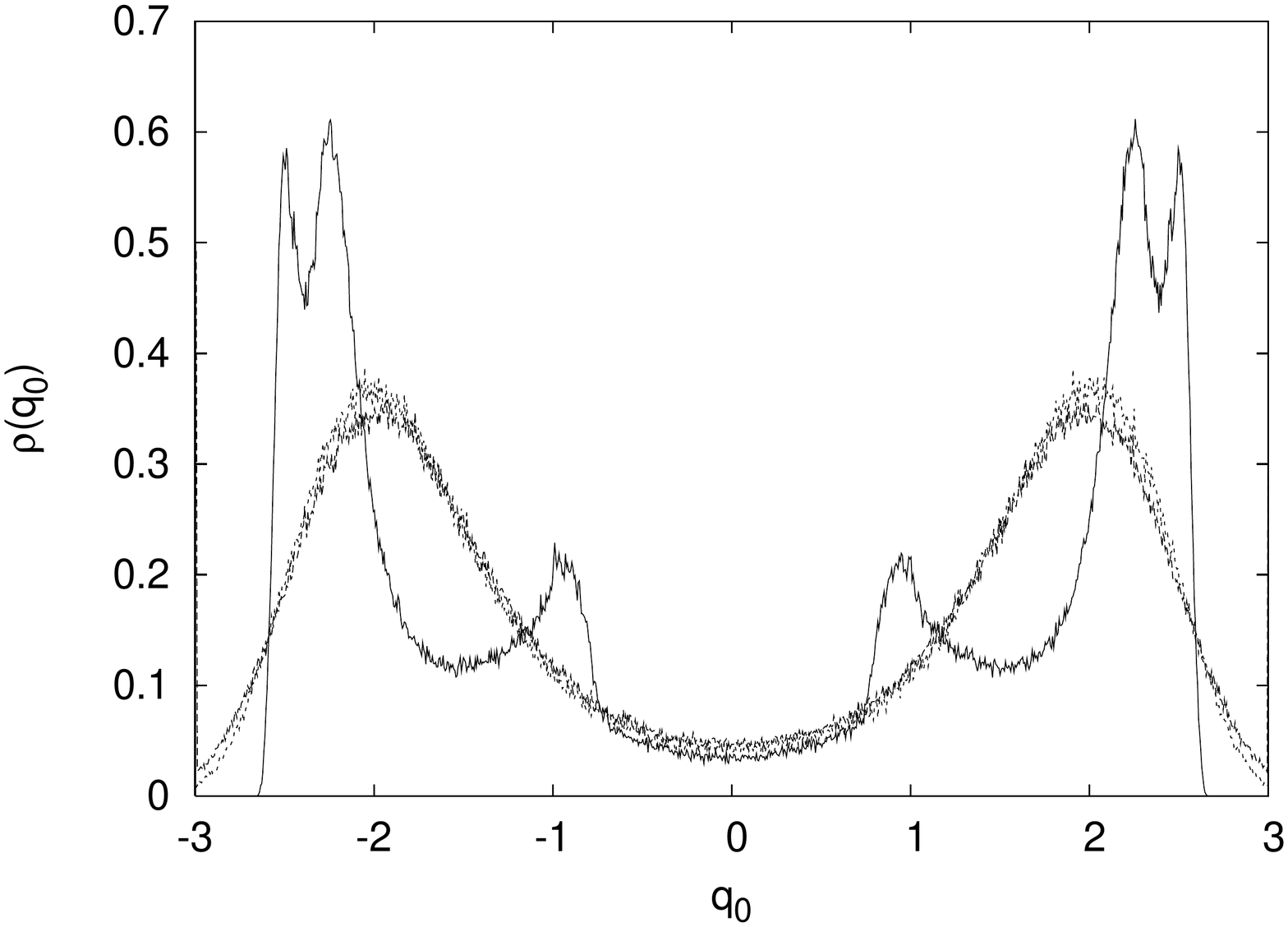}
  \end{center}
  \caption{\label{fig:dwallq0pdf} 
PDFs of the slow variable $q_0$ for the DW model with $\widetilde{\epsilon} = 1$, 
  i.e. $\epsilon \sim 0.1$ (full line), 
  the WNDW model (dashed line) and the DW model with $\widetilde{\epsilon} = 10^{-2}$, 
  i.e. $\epsilon \sim 10^{-3}$ (dotted line). 
  In the limit $\epsilon \to 0$, the PDFs of the deterministic model and of the stochastic model collapse.
}
\end{figure}

Let us notice that, in this case, because of the skew structure of the original system, the stochastic 
modeling is (relatively) simple and, differently from the generic case, the noise is additive.  
The time signal ${q_0}(t)$ obtained from the WNWD model is reported in 
Fig. \ref{fig:dwwnoiseq0}. One observes strong similarities in the long-time transition 
statistics with respect to the deterministic model, even though the PDFs of the slow variable are quite different from one another, see Fig. \ref{fig:dwallq0pdf}.   
  


\begin{figure}[htb]
  \vspace*{2mm}
  \begin{center}
    \includegraphics[width=0.9\textwidth]{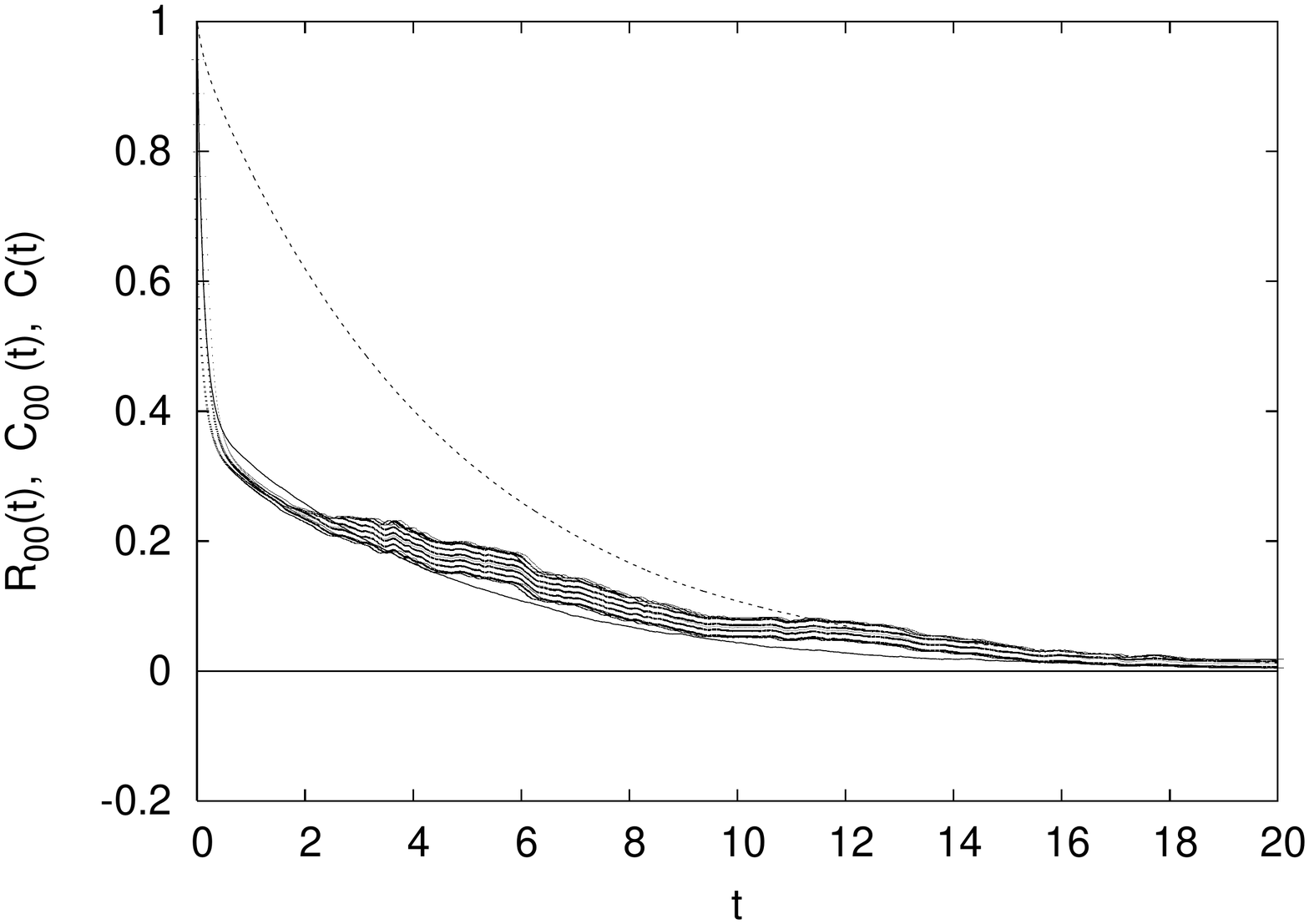}
  \end{center}
  \caption{\label{fig:dwwnoisefrq0} 
WNDW model: autocorrelation $C_{00}(t)$ (dashed line), self-response $R_{00}(t)$, 
  with statistical error bars, and the correlation function 
  $C(t)$ predicted by the FRR (full line).
}
\end{figure}

The FR properties of the WNDW model are reported in Fig. \ref{fig:dwwnoisefrq0}. The slow variable is 
distributed according to $\sim e^{-V(q_0)/K}$, with $K=\sigma^2/2$, and the FR theorem 
prediction is verified, i.e. one has a good agreement between $R_{00}(t)$ and the  
correlation function $C(t)$.

We redefine, as already seen 
when discussing the stochastic model approximating the Lorenz-96 system, 
the quadratic cross-response function $R_{sf}^{(q)}(t)$  
as the root mean square growth of the error $\delta q_0(t)$ induced by two different noise realizations. 

In Fig. \ref{fig:dwallCRF}, the behavior of $R_{sf}^{(q)}(t)$ for the deterministic DW system and its 
stochastic model 
is reported. The WNDW model is not able to 
reproduce the two-time behavior of the deterministic model, mainly due to the impossibility to control the 
amplitude of the initial perturbation. 
Because of that, the error on the climatic state of the system saturate very quickly, as soon as the trajectory starts 
jumping between the wells.


\section{Discussion and conclusive remarks}
\label{sec:conclusions}

In this paper we have presented a detailed investigation of the Fluctuation-Response properties 
of chaotic systems with fast and slow dynamics. 
The numerical study has been performed on two models, namely the 360-variable Lorenz-96 system, 
with reciprocal feedback between fast and slow variables, and a simplified low dimensional system, 
both of which are able to capture the main features, and related difficulties, typical of the multiscale 
systems. 
The first point we wish to emphasize is how, even in non Hamiltonian systems, a generalized Fluctuation-Response 
Relation (FRR) holds. This allows for a link between the average relaxation of perturbations and the statistical 
properties (correlation functions) of the unperturbed system. Although one has non Gaussian statistics, the 
correlation functions of the slow (fast) variables have at least a qualitative resemblance with the response 
functions to perturbations on the slow (fast) degrees of freedom. 
The average response function of a slow variable to perturbations of the fast degrees of freedom is zero, 
nevertheless the impact of the fast dynamics on the slowly varying components cannot be neglected. 
This fact is clearly highlighted by the behavior of a suitable quadratic response function. 
Such a phenomenon, which can be regarded as a sort of sensitivity of the slow variables to variations of the 
fast components, has an important consequence for the modeling of the slow dynamics in terms of a Langevin 
equation. 
Even an optimal model (i.e. able to mimic autocorrelation and self-response of the slow variable), beyond 
a certain intrinsic time interval, can give just statistical predictions, in the sense that, at most, one 
can hope to have an agreement among the statistical features of system and model.  
In stochastic dynamical systems, one has to deal with a similar behavior: the relevant `complexity' of the 
systems is obtained  by considering the divergence of nearby trajectories evolving under two different 
noise realizations. Therefore a good model for the slow dynamics (e.g. a Langevin equation) must show a sensitivity 
to the noise.

\section{Appendix A: Generalized FRR}
\label{sec:AppendixA}

In this Appendix  we give a derivation, under general rather hypothesis,
of a generalized FRR.
Consider a dynamical system $ {\bf x}(0) \to {\bf x}(t)=U^t {\bf x}(0)$
with states ${\bf x}$ belonging to a $N$-dimensional vector space.  
For the sake of generality, we will
consider the case in which the time evolution can also be not
completely deterministic (\textit{e.g.} stochastic differential equations). 
We assume the existence of an invariant probability distribution
$\rho({\bf x})$, for which some ``absolute continuity'' type conditions
are required (see later), 
and the mixing character of the system (from which its
ergodicity follows).  Note that no assumption is made on $N$.

Our aim is to express the average response of a generic observable $A$
to a perturbation, in terms of suitable correlation functions,
computed according to the invariant measure of the unperturbed system.
At the first step we study the behavior of one component of ${\bf  x}$,
 say $x_i$, when the system, described by $\rho({\bf x})$, is
subjected to an initial (non-random) perturbation such that 
${\bf  x}(0) \to {\bf x}(0) + \Delta {\bf x}_{0}$.  
  This
instantaneous kick\footnote{The study of an
  ``impulsive'' perturbation is not a severe limitation, e.g. in
  the linear regime from the (differential) linear response one
  understands the effect of a generic perturbation.} 
modifies the density of the system into
$\rho'({\bf x})$, related to the invariant
distribution by $\rho' ({\bf x}) = \rho ({\bf x} - \Delta {\bf x}_0)$.
We introduce the probability of transition from ${\bf x}_0$ at time
$0$ to ${\bf x}$ at time $t$, $W ({\bf x}_0,0 \to {\bf x},t)$. For a
deterministic system, with evolution law $ {\bf x}(t)=U^{t}{\bf
  x}(0)$, the probability of transition reduces to $W ({\bf x}_0,0 \to
{\bf x},t)=\delta({\bf x}-U^{t}{\bf x}_{0})$, where $\delta(\cdot)$ is the
Dirac's delta.  Then we can write an expression for the mean value of
the variable $x_i$, computed with the density of the perturbed system:
\begin{equation}
\label{A.1}
\Bigl \langle x_i(t) \Bigr \rangle ' = 
\int\!\int x_i \rho' ({\bf x}_0) 
W ({\bf x}_0,0 \to {\bf x},t) \, d{\bf x} \, d{\bf x}_0  \; .
\end{equation}
The mean value of $x_i$ during the unperturbed evolution can be written in
a similar way:
\begin{equation}
\label{A.2}
\Bigl \langle x_i(t) \Bigr \rangle = 
\int\!\int x_i \rho ({\bf x}_0) 
W ({\bf x}_0,0 \to {\bf x},t) \, d{\bf x} \, d{\bf x}_0  \; .
\end{equation}
Therefore, defining $\overline{\delta x_i} =  \langle x_i \rangle' -
\langle x_i \rangle$, we have:
\begin{eqnarray}
\label{A.3}
\overline{\delta x_i} \, (t)  &=&
\int  \int x_i \;
F({\bf x}_0,\Delta {\bf x}_0) \;
\rho ({\bf x}_0) W ({\bf x_0},0 \to {\bf x},t) 
\, d{\bf x} \, d{\bf x}_0 \nonumber \\
&=& \Bigl \langle x_i(t) \;  F({\bf x}_0,\Delta {\bf x}_0) \Bigr \rangle
\end{eqnarray}
where
\begin{equation}
\label{A.4}
F({\bf x}_0,\Delta {\bf x}_0) =
\left[ \frac{\rho ({\bf x}_0 - \Delta {\bf x}_0) - \rho ({\bf x}_0)}
{\rho ({\bf x}_0)} \right] \; .
\end{equation}
Let us note here that the mixing property of the system is required so
that the decay to zero of the time-correlation functions assures the
switching off of the deviations from equilibrium.
 
For an infinitesimal perturbation $\delta {\bf x}(0) = (\delta x_1(0)
\cdots \delta x_N(0))$, if $\rho({\bf x})$ is non-vanishing and
differentiable, the function in (\ref{A.4}) can be expanded to first
order and one obtains:
\begin{eqnarray}
\label{A.5}
\overline{\delta x_i} \, (t)  &=&
- \sum_j \Biggl
\langle x_i(t) \left. \frac{\partial \ln \rho({\bf x})}{\partial x_j} 
\right|_{t=0}  \Biggr \rangle \delta x_j(0) \nonumber \\
&\equiv&
\sum_j R_{ij}(t) \delta x_j(0)
\end{eqnarray}
which defines the linear response 
\begin{equation}
\label{A.6}
R_{ij}(t) = - \Biggl \langle x_i(t) \left.
 \frac{\partial \ln \rho({\bf x})} {\partial x_j} \right|_{t=0}
\Biggr  \rangle 
\end{equation} 
of the variable $x_i$ with respect to a perturbation of $x_j$.
One can easily repeat the computation for a generic observable
$A({\bf x})$:
\begin{equation}
\label{A.7}
\overline{\delta A} \, (t)  = -\sum_j
\Biggl \langle A({\bf x}(t)) \left.
\frac{\partial \ln \rho({\bf x})} {\partial x_j} \right|_{t=0} 
\Biggr \rangle \delta x_j(0) \,\, .
\end{equation}

For Langevin equations, the differentiability of $\rho({\bf X})$ is 
well established. On the contrary, 
one could argue that in a chaotic deterministic
dissipative system the above machinery cannot be applied, because the
invariant measure is not smooth at all. 
Typically the invariant measure of a chaotic attractor 
 has a multifractal character and its  Renyi dimensions $d_q$ are
not constant (Paladin and Vulpiani, 1987). 
In chaotic dissipative systems the invariant measure is singular,
however  the previous derivation of the FRR is  still valid 
if one considers perturbations along the expanding directions.  For a
mathematically oriented presentation see Ruelle (1998).
A general response function has two contributions, corresponding respectively
to the expanding (unstable) and the contracting (stable) directions of
the dynamics. The first contribution can be associated to some
correlation function of the dynamics on the attractor (i.e. the
unperturbed system). On the contrary this is not true for the second
contribution (from the contracting directions), this
part to the response is very difficult to extract numerically
(Cessac and Sepulchre, 2007). 
In  chaotic deterministic systems, in order to have a differentiable 
invariant measure, one has to invoke the stochastic regularization 
(Zeeman 1990). If such a method is not feasible, one can use the direct approach 
by Abramov and Majda (2007). 
For a study of the  FRR in chaotic atmospheric systems, see
Dymnikov and Gritsoun (2005)  and Gritsoun and Branstator (2007).

Let us notice that a small amount
of noise, that is always present in a physical system, smoothen the
$\rho({\bf x})$ and the FRR can be derived.  We recall that
this ``beneficial'' noise has the important role of selecting the
natural measure, and, in the numerical
experiments, it is provided by the round-off errors of the
computer. We stress that the assumption on the smoothness of the
invariant measure allows to avoid subtle technical difficulties.

\section{Appendix B: A general remark on the decay of correlation functions}
\label{sec:AppendixB}

Using some general arguments  one has that all the (typical) correlation
functions at large time delay have to relax to zero with the same
characteristic time, related to spectral properties of the  operator
$\hat{\bf L}$ which rules the time evolution of the $P({\bf X},t)$:
\begin{equation} 
\label{B.1}
{ \partial \over {\partial t}} P({\bf X},t)=\hat{\bf L} P({\bf X},t) \, .
\end{equation}
In the case of ordinary differential equations 
\begin{equation} 
\label{B.2}
dX_i/dt=Q_i({\bf X}) \,\,\,\,\ i=1,\cdots,N
\end{equation}
the operator $\hat{\bf L}$ has the shape
\begin{equation} 
\label{B.3}
\hat{\bf L} P({\bf X},t)
= -\sum_i {\partial \over {\partial X_i}}
\Bigl( Q_i({\bf X})P({\bf X},t)\Bigr)\, .
\end{equation}
For Langevin equations i.e. in (\ref{B.2}) 
$Q_i$ is replaced by $ Q_i+{\eta}_i$ where $\{ {\eta}_i \}$ 
are Gaussian processes with $<\eta_i(t)>=0$ and  
$<{\eta}_i(t){\eta}_j(t')>=2\Lambda_{i,j}\delta(t-t')$, 
 one has
\begin{equation} 
\hat{\bf L} P({\bf X},t)
= -\sum_i {\partial \over {\partial X_i}}
\Bigl( Q_i({\bf X})P({\bf X},t)\Bigr) 
+ \sum_{ab} \Lambda_{i,j} 
{\partial^2 \over {\partial X_i \partial X_i} } P({\bf X},t)  \, .
\end{equation}

Let us introduce the eigenvalues $\{\alpha_k\}$ and the eigenfunctions
$\{\psi_k\}$ of ${\cal L}$:
\begin{equation} 
\label{B.4}
\hat{L}\psi_k=\alpha_k \psi_k \, .
\end{equation}
Of course $\psi_0=P_{inv}$ and $\alpha_0=0$, and typically in mixing
systems $Re\, \alpha_k <0$ for $k=1,2,...$.
Furthermore assuming that coefficient $\{g_1,g_2,...\}$ and 
$\{h_1,h_2,...\}$ exist such that functions $g({\bf X})$ and 
$h({\bf X})$ are uniquely expanded as
\begin{equation} 
\label{B.5}
g({\bf X})=\sum_{k=0}g_k \psi_k({\bf X}) \,\,\, , \,\,\,
h({\bf X})=\sum_{k=0}h_k \psi_k({\bf X}) \,\, ,
\end{equation}
so we have
\begin{equation} 
\label{B.6}
C_{g,f}(t)= \sum_{k=1} g_k h_k <\psi_k^2> e^{\alpha_k t} \, ,
\end{equation}
where $C_{g,f}(t)= <g({\bf X}(t)) h({\bf X}(t))> - 
<g({\bf X})>< h({\bf X})>$.
For ``generic'' functions $g$ and $f$, i.e. if they are not orthogonal
to $\psi_1$ so that $g_1\neq 0$ and $h_1\neq 0$, at large time the 
correlation $C_{g,f}(t)$ approaches to zero as
\begin{equation} 
\label{B.7}
C_{g,f}(t) \sim e^{-t/\tau_c} \,\,\, , \,\,\, 
\tau_c= {1 \over {|Re \, \alpha_1|}} \, .
\end{equation}

In some cases, e.g. very intermittent systems like the Lorenz model at $r \simeq 166.07$, 
$Re \, \alpha_1 = 0$ so the decay is not exponentially fast.

\section{Appendix C: Lyapunov exponent in dynamical systems with noise}
\label{sec:AppendixC}

In systems with noise, the simplest  way to introduce
the Lyapunov exponent  is to
treat the random term as a time-dependent term.
Basically one  considers
the separation of two close trajectories with the same realization of
noise. 
Only for sake of simplicity consider a one-dimensional Langevin
 equation 
\begin{equation}
\label{C.1}
{{d x} \over {d t} }= - { {\partial V(x)} \over {\partial x}} + \sigma \, \eta \, ,
\end{equation} 
where $\eta(t)$ is a white
noise and $V(x)$ diverges for $\mid x\mid \to \infty$, like, e.g., the
usual double well potential $V=-x^2/2+x^4/4$.
 
The Lyapunov exponent $\lambda_{\sigma}$, associated with the
separation rate of two nearby trajectories with the same realization
of $\eta(t)$, is defined as 
\begin{equation}
\label{C.2}
\lambda_{\sigma}=\lim_{t\to \infty} { 1 \over t} \ln |z(t)|
\end{equation} 
where the evolution of the tangent
vector is given by: 
\begin{equation} 
\label{C.3}
{ {d z} \over {d t} }= -
 {{\partial^2 V(x(t))} \over {\partial x^2}}z(t).
 \end{equation} 
The quantity  $\lambda_{\sigma}$ obtained in the previous way, although
well defined, i.e. the Oseledec theorem (Bohr et al., 1998) holds, it is not always a
useful characterization of complexity. 

Since
the system is ergodic with invariant probability distribution $P(x)=C_1
e^{- V(x)/C_2}$, where $C_1$ is a normalization constant and $C_2=\sigma^2/2$, one has: 
\begin{equation} 
\label{C.4}
\begin{array}{ll}
\lambda_{\sigma}&=\lim_{t\to \infty} { 1 \over t} \ln |z(t)| 
= -\lim_{t\to \infty} { 1 \over t} \int_0^t \partial^2_{xx} V(x(t'))
dt' \\
& \\
&= -C_1 \int \partial^2_{xx} V(x) e^{-V(x)/C_2} \,\, dx 
= -{C_1 \over C_2 } \int (\partial_x V(x))^2 e^{-V(x)/C_2} \,\, dx <
0 \, .  
\end{array} 
\end{equation} 
This  has a rather intuitive
meaning: the trajectory $x(t)$ spends most of the time in one of the
``valleys'' where $-\partial^2_{xx} V(x) < 0$ and only  short intervals
on the ``hills'' where $-\partial^2_{xx} V(x) > 0$, so that the distance
between two trajectories evolving with the same noise realization
decreases on average.  
 The previous  result for the $1D$ Langevin equation can easily be
generalized to any dimension for gradient systems if the noise is
small enough (Loreto et al., 1996).

A negative value of $\lambda_{\sigma}$ 
implies a fully predictable process only if the
realization of the noise is known.  In the case of two initially close
trajectories evolving under two different noise realizations, after a
certain time $T_{\sigma}$, the two trajectories can be very distant,
because they can be in two different valleys. For $\sigma \to 0$, due
to the Kramers formula (Gardiner, 1990), 
one has $T_\sigma \sim e^{\Delta
V/\sigma^2}$, where $\Delta V$ is the difference between the values of
$V$ on the top of the hill and at the bottom of the valley.

Let us now discuss the main difficulties in defining the notion 
of `complexity' of an evolution law with a random perturbation,
discussing a simple case.
Consider the $1D$ map 
\begin{equation}
\label{C.5}
 x(t+1)=f[x(t),t]+\sigma w(t), 
\end{equation}
where $t$ is an integer and $w(t)$ is an uncorrelated random process,
e.g. $w$ are independent random variables uniformly distributed in
$[-1/2,1/2]$. For the largest LE $\lambda_{\sigma}$, as defined in
(\ref{C.2}), now one has to study the equation
\begin{equation}
\label{C.6}
 z(t+1)=f'[x(t),t] \,z(t), 
\end{equation} 
where $f'=df/dx$.
 
Following the approach in (Paladin et al., 1995) 
let $x(t)$ be the trajectory starting at
$x(0)$ and $x'(t)$ be the trajectory starting from $x'(0)=x(0)+\delta
x (0)$.  Let $\delta_0 \equiv |\delta x(0)|$ and indicate by $\tau_1$
the minimum time such that $|x'(\tau_1)-x(\tau_1)|\ge \Delta$. Then,
we put $x'(\tau_1)=x(\tau_1)+\delta x(0)$ and define $\tau_2$ as the
time such that $|x'(\tau_1+\tau_2)-x(\tau_1+\tau_2)|>\Delta$ for the
first time, and so on. In this way the Lyapunov exponent 
can be defined  as
\begin{equation} 
\lambda= 
{1 \over \overline{\tau} } \,
\ln \left( { \Delta \over \delta_0} \right)\,
\end{equation}
being $\overline{\tau}=\sum \tau_i/{\cal N}$ where ${\cal N}$ is the number
of the intervals in the sequence.
If the above procedure is applied 
by considering the same noise realization for both trajectories, 
$\lambda$ in (\ref{C.2}) coincides with $\lambda_{\sigma}$ 
(if $\lambda_{\sigma}>0$). Differently, by considering two
different realizations of the noise for the two trajectories, we have a 
new quantity 
\begin{equation} 
K_{\sigma}= \, {1 \over \overline {\tau} } \,
 \ln \left( { \Delta \over \delta_0}\right)\,,
\end{equation} 
which  naturally arises in the framework of
information theory and algorithmic complexity theory: note that
$K_{\sigma}/\ln 2$ is the number of bits {\it per}
unit time one has to specify in order to transmit the sequence 
with a precision $\delta_0$,
The generalization of the above treatment  to $N$-dimensional maps or 
to ordinary differential equations is straightforward.

If the fluctuations of the effective Lyapunov exponent $\gamma(t)$ 
(in the case of (\ref{C.5}) $\gamma(t)$ is nothing but  
$\ln |f'(x(t))|$)
are very small (i.e. weak intermittency) one has 
$K_{\sigma} \, = \, \lambda +O(\sigma/\Delta)\,.$

The interesting situation happens for strong intermittency when there
are alternations of positive and negative $\gamma$ during long time
intervals: this induces a dramatic change for the value of
$K_{\sigma}$.  
Numerical  results on intermittent maps (Paladin et al., 1995)
 show that the same system 
can be regarded either
as regular (i.e. $\lambda_{\sigma}<0$), when the same noise
realization is considered for two nearby trajectories, or as chaotic
(i.e. $K_{\sigma}>0$), when two different noise realizations are
considered. We can say that a negative $\lambda_{\sigma}$ for some
value of $\sigma$ in not an indication that ``noise induces order'';
a correct conclusion is that noise can induce synchronization.

\begin{acknowledgements} 
We warmly thank A. Mazzino, S. Musacchio, R. Pasmanter and A. Puglisi for interesting discussions and suggestions, 
and two anonymous referees for their contructive criticism in reviewing this paper. 
\end{acknowledgements}




\newpage

\end{document}